%% file: main.tex
\documentclass{aa}

\usepackage{xcolor}

\usepackage{soul}
\usepackage{natbib}
\usepackage{amsmath}
\usepackage{bm}
\usepackage{caption}
\usepackage{subcaption}
\usepackage{hyperref}
\usepackage{graphicx}
\usepackage{natbib}
\usepackage{url}
\usepackage{ulem}
\usepackage{xspace}
\usepackage{array}
\usepackage{placeins}

\bibpunct{(}{)}{;}{a}{}{,}

\hypersetup{
    colorlinks=true,
    linkcolor=blue,
    citecolor=blue,
    filecolor=magenta,
    urlcolor=cyan,
    }

\def \cesamxx{Cesam2k20\xspace}

\newcommand{\anna}[1]{{ #1}}

\newcommand{\lm}[1]{{ #1}}

\newcommand{\Ekma}{\operatorname{{E\kern-.08em k}}} 
\newcommand{\Nuss}{\operatorname{{N\kern-.09em u}}} 
\newcommand{\Pram}{\operatorname{{P\kern-.03em m}}} 
\newcommand{\Pran}{\operatorname{{P\kern-.03em r}}} 
\newcommand{\Rayl}{\operatorname{{R\kern-.04em a}}} 
\newcommand{\Reyn}{\operatorname{{R\kern-.04em e}}} 
\newcommand{\Ross}{\operatorname{{R\kern-.04em o}}} 

\newcommand{\diff}[2]{\frac{{\rm d} #1}{{\rm d} #2}}

\newcolumntype{L}[1]{>{\raggedright\let\newline\\\arraybackslash\hspace{0pt}}m{#1}}

\graphicspath{{./}{images/}}

\begin{document}

\title{Radial differential rotation leading to dipole collapse in pre-main-sequence stars}
\titlerunning{Dipole collapse in PMS stars}

\author{A.~Guseva\inst{1,2} 
\and L.~Manchon\inst{2} 
\and L.~Petitdemange\inst{2}
\and C.~Pinçon\inst{3,2}}

\institute{Department of Fluid Mechanics, Universitat Politècnica de Catalunya (UPC), 08019 Barcelona, Spain \\
Corresponding author: anna.guseva@upc.edu \and
LIRA, Observatoire de Paris, Université PSL, Sorbonne Université, Université Paris cité, CY Cergy Paris Université, CNRS, 75014 Paris, France
\and Institut d'Astrophysique Spatiale, Université Paris-Saclay, Orsay, France}

\date{Received XXX / Accepted XXX}

\abstract{Despite significant progress in the observational characterization of stellar magnetic fields, the physical processes that govern their intensity and topology, which could certainly result from their formation history, remain poorly understood. 
During the pre-main sequence (PMS) phase, the inner layers tend to contract and a radiative core gradually develops. In contrast, the convective envelope is gradually braked through the magnetic interactions with the accretion disk and winds, thus slowly developing differential rotation inside the star. It is likely during this PMS phase that the dynamo processes that efficiently generated strong dipolar magnetic fields through vigorous convective motions in protostars become highly perturbed, leading to the observed diversity in the magnetism on the main sequence (MS).}
{We aim to study the stability of dipolar magnetic fields inherited from the proto-stellar phase by considering the emergence of a large-scale radial differential rotation resulting from the combined actions of contraction and of the interactions with the surrounding medium.}
{We perform three-dimensional convective dynamo simulations of rotating spherical shells with an imposed differential rotation (shear) between the bottom and top boundaries. We use the anelastic approximation that allows us to consider background density and gravity profiles and convective zone thicknesses close to those predicted in PMS low-mass stars by the one-dimensional stellar evolution code \cesamxx.
We then carry out a parameter study by systematically varying the shear amplitude.
}
{Radial differential rotation can induce dipole collapse leading to weaker and oscillatory magnetic fields. We highlight that the stability of dipolar dynamos mainly depends on the relative importance of 
shear measured by a shear-based Rossby number compared to the vigor of convective motions as measured by the convective Rossby number.    We show that the stability criterion depends on the field strength and the size of the radiative core. Differential rotation seems to perturb $\alpha^2$ dynamo mechanism responsible for dipolar magnetic fields by shearing poloidal field lines and by affecting turbulent magnetic transport processes.}
{PMS phase can represent a critical period for the magnetic properties of stars as the development of shear layers can perturb the stability of strong initial dipoles. Applying the stability criterion in PMS stellar evolution models,
we qualitatively reproduce the trends observed in the magnetic topologies of low-mass stars when assuming an efficient internal angular momentum redistribution process. This
suggests that stellar magnetic properties are intimately related to the PMS angular momentum evolution.
}


\maketitle

\section{Introduction}

\begin{figure}[t]
    \includegraphics[width=0.98\linewidth]{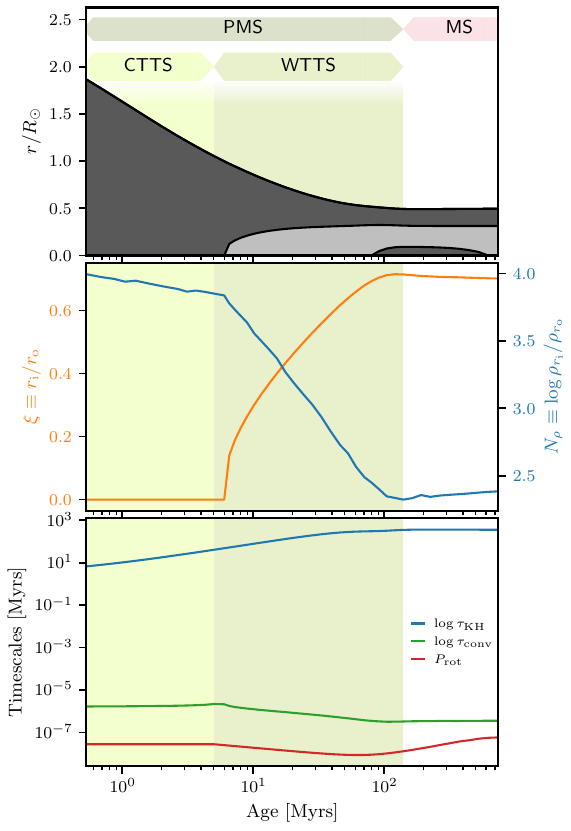}
    \caption{Evolution of a $0.55M_\odot$ model. Top: Kippenhahn diagram. Radius of interfaces in solar unit as a function of the model's age. Convective zones (resp. radiative zones) are represented as dark (resp. light) grey areas. Middle: Aspect ratio (left axis) and density contrast (right axis). Inner radius $r_{\rm i}$ is the bottom radius of the convective envelope, while the outer radius $r_{\rm o}$ is $90\%$ of the photospheric radius. Bottom: Kelvin-Helmholtz (blue) and convection (green) timescales and surface rotation period (red). The shaded areas correspond to the classical T Tauri phase (CTTS; light green) which coincides in our case with the disc-locking phase, and weak T Tauri phase (WTTS; green).}
    \label{fig:kippenhahn}
\end{figure}

\input{intro_charly}

\section{Main structural properties of PMS stars}
\label{section:early_stell_evol}
In this section, we briefly remind the global internal properties of PMS stars using predictions from one-dimensional evolutionary models. This introductory part will serve as a guide to justify the assumptions used in our 3D numerical dynamo simulations presented subsequently. 

\begin{figure}[t]
    \includegraphics[width=0.98\linewidth]{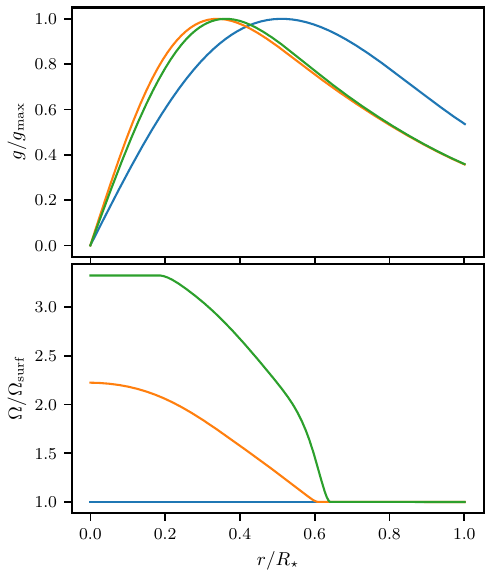}
    \caption{Normalized gravitational acceleration $g/g_{\rm max}$ (top panel) and normalized angular velocity $\Omega/\Omega_{\rm surf}$ (bottom panel) as a function of the radius. These profiles are shown at the beginning (blue), middle (orange) and end (green) of the WTTS phase. Same model as in Fig.~\ref{fig:kippenhahn}.}
    \label{fig:grav_omega}
\end{figure}

\subsection{Stellar models with \cesamxx}
\label{subsection:cesam}
The evolutionary tracks of stellar structure models are computed with open-source software \cesamxx \citep{manchon2025cesam2k20,Morel1997,Morel2008,Marques2013}. The microphysics used to compute these models is  summarized in App. \ref{appendix:cesam_phy}. All the models presented in this paper neglect the transport of chemical elements which is a fair approximation because of the short PMS time compare to the chemical transport timescale. The transport of angular momentum inside radiative zones includes the effect of the meridional circulation and of the shear-induced turbulence. Convective zones are roughly supposed to rotate as solid body owing to the simplified hypothesis of very efficient convective mixing. The initial distribution of angular momentum inside the star is set with the disc-locking scenario of \citet{Bouvier1997}: the convective zones rotate rigidly at fixed period of the disk $P_{\rm disk}$, during its lifetime $\tau_{\rm disc}$. The lifetime of this disk is not a prediction of our model, but an input. The reference model discussed in this illustrative section has $\tau_{\rm disk}=5$ Myrs, $P_{\rm disk} = 10$ days and a mass $M=0.55~M_\odot$.

\begin{figure*}[t]
     \centering
     \begin{subfigure}[b]{0.48\textwidth}
         \centering
         \includegraphics[width=\textwidth]{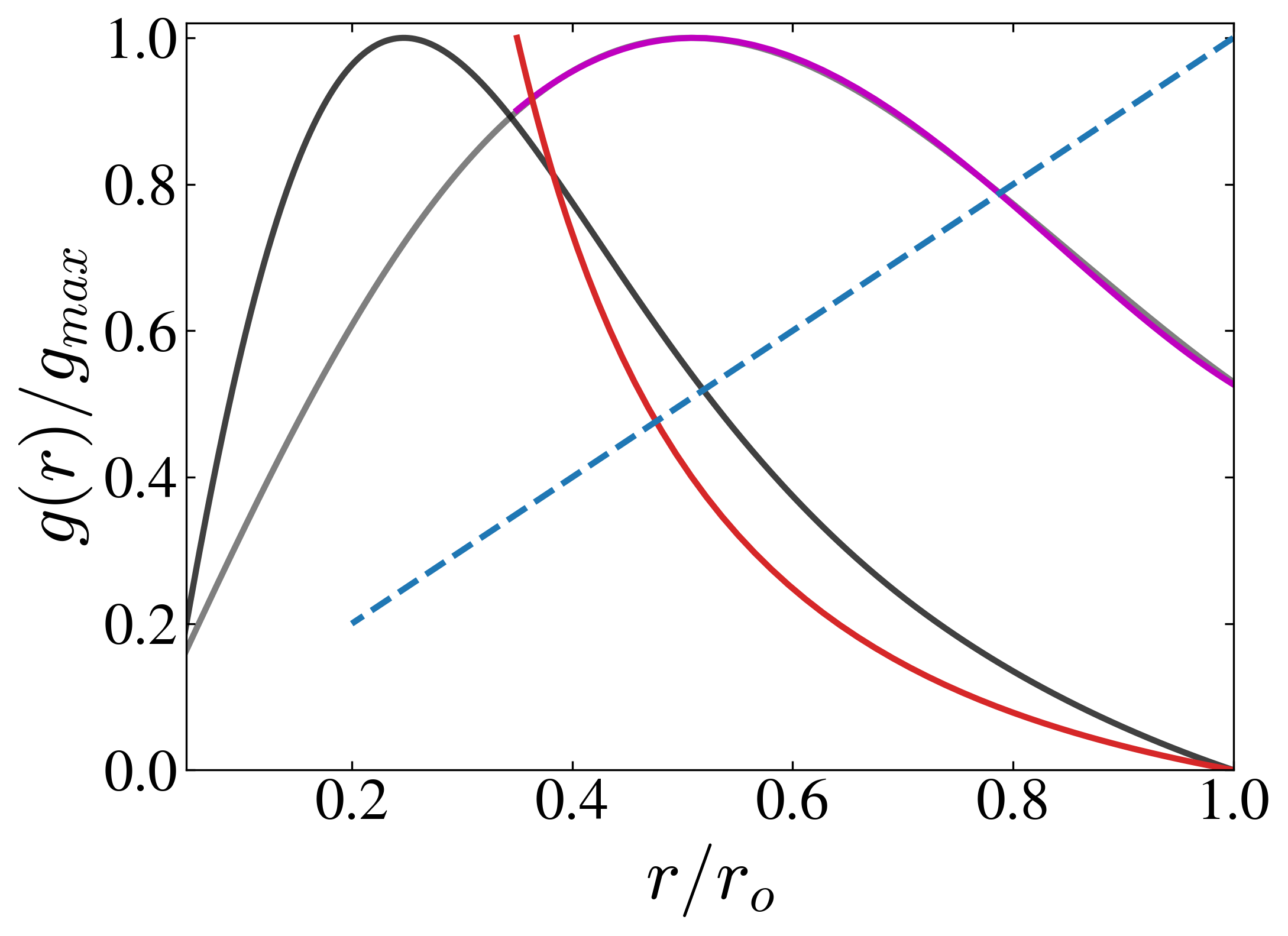}
         \caption{}
     \end{subfigure}
     \hfill
          \begin{subfigure}[b]{0.48\textwidth}
         \centering
         \includegraphics[width=\textwidth]{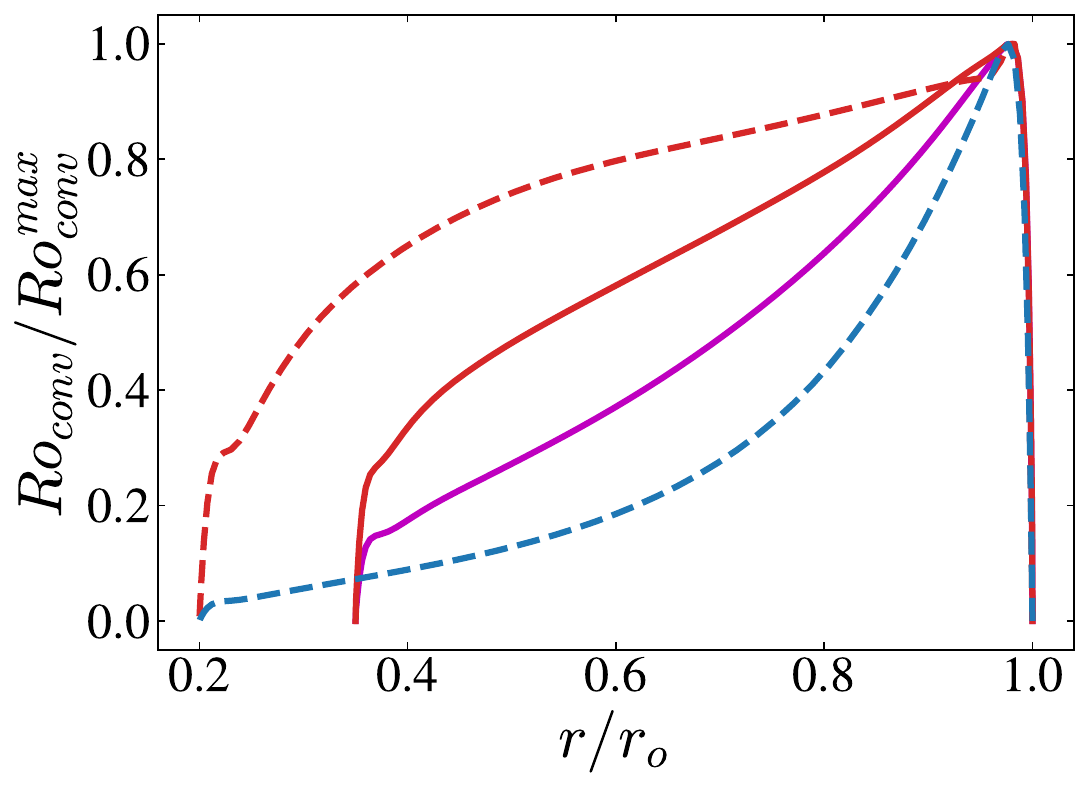}
         \caption{}
     \end{subfigure}
     \vfill
        \begin{subfigure}[b]{0.29\textwidth}
         \centering
         \includegraphics[width=\textwidth]{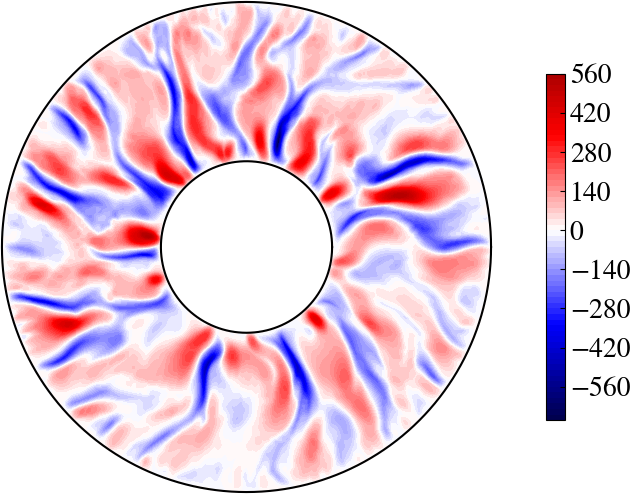}
         \caption{}
          \end{subfigure}
         \hfill
        \begin{subfigure}[b]{0.29\textwidth}
         \centering
         \includegraphics[width=\textwidth]{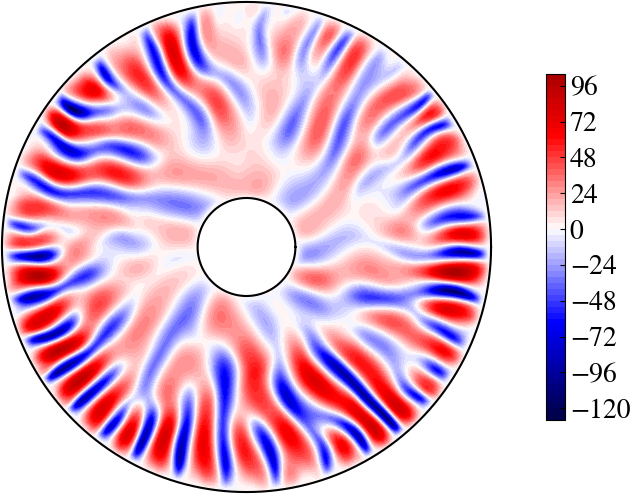}
         \caption{}
          \end{subfigure}
                 \hfill
        \begin{subfigure}[b]{0.29\textwidth}
         \centering
         \includegraphics[width=\textwidth]{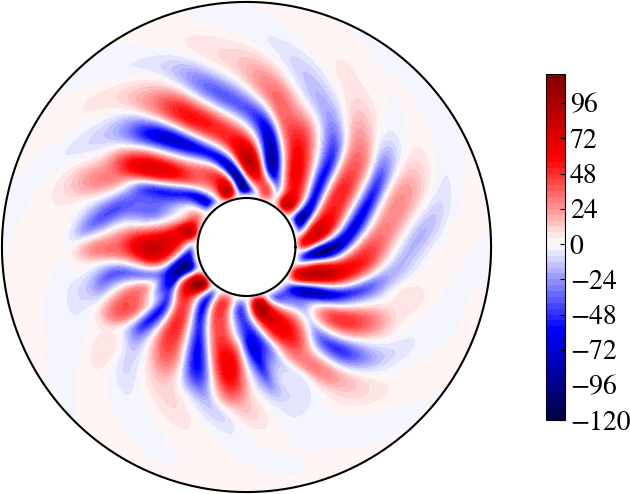}
         \caption{}
          \end{subfigure}
     \caption{(a) Normalized gravity profiles from stellar evolution code \cesamxx for a star of $0.8 M_\odot$ at an early stage of stellar evolution ($1$ Myr, in light gray) and at the MS ($4.9 \cdot 10^3$ Myrs, dark gray). Example of imposed gravity profiles in DNS: $g\propto 1/r^2$ for radii ratio $\xi = 0.35$ (in red), $g\propto r$ and $\xi = 0.2$ (in dashed light blue) and a typical profile from \cesamxx with $\xi=0.35$ (in magenta, see Eq.~\eqref{eq:g_r_cesam} for detail). (b) Local convective Rossby number, normalized with its maximum, as a function of $r/r_{\rm o}$ for the runs \texttt{gr2\_2}, \texttt{gr\_1}, \texttt{gc\_2}, associated with the gravity profiles of panel (a) (same color code), and for the run \texttt{gr2\_3}with $g \propto 1/r^2$ and $\xi = 0.2$ (in dashed red). See details of the runs in Table~\ref{tab:sim_param}. (c) Equatorial cut of instantaneous radial velocity for $g\propto 1/r^2$, $\xi = 0.35$ (run  \texttt{gr2\_2}). (d) Same for $g\propto r$, $\xi =0.2$ (run \texttt{gr\_1}) (e) Same for  $g \propto 1/r^2$, $\xi = 0.2$ (run \texttt{gr2\_3}). \anna{Velocities in panels (c-e) were normalized with $\nu/(r_o - r_i)$, i.e. kinematic viscosity $\nu$  and  the thickness of convective zone.}
     }
    \label{fig:cesam}
\end{figure*}

\subsection{Constraints from stellar evolution}

\cesamxx starts the evolution during the T-Tauri phase, when the star has already accreted all its mass and is still surrounded by a debris disk. Such a star is initially fully convective, as illustrated on the Kippenhahn diagram a $0.55M_\odot$ model in Fig. \ref{fig:kippenhahn}, top panel. The T-Tauri phase, which we assimilate here to the PMS phase, is decomposed into two stages: the classical T-Tauri (CTTS) phase where the star is surrounded by an optically thick debris disk and the weak T-Tauri (WTTS) phase when the disk has become optically thin and cease to influence the star. CTTS rotate fast and are fully or almost fully convective; at this stage, an intense magnetic field is assumed to lock the convective zone rotation to the disc rotation (disk-locking). Once the disk dissipates, the angular velocity of the convective zones is controlled by the star contraction and wind braking.

Fig. \ref{fig:grav_omega} displays normalized gravity and angular velocity $\Omega$ profiles for three stages: end of CTTS phase (blue), and middle (orange) and end (green) of WTTS phase. For our $0.55M_\odot$ model, the CTTS phase ends while the model is still fully convective and thus rotates as a solid body. When the star decouples from the disc, and the radiative core appears, the radial differential rotation inside the radiative zone increases with time. The internal increase is predominantly controlled by contraction, despite the presence \lm{of shear-induced mixing and of meridional circulation that tend to smooth its effect}. We note that a convective core temporarily formed at the end of WTTS phase where (Fig.~\ref{fig:kippenhahn}, top panel), also visible through the flat part of the $\Omega$-profile in Fig.~\ref{fig:grav_omega} (green)\footnote{\anna{The short appearance of a convective core,  preceding ignition of H1 burning, corresponds to the first steps of CNO cycle. The chain of reactions, corresponding to ${}^{14}{\rm N(p},\gamma){}^{15}{\rm O}$ 
${}^{12}{\rm C(p},\gamma){}^{13}{\rm N}(,\beta^+ \nu){}^{13}{\rm C(p},\gamma){}^{14}{\rm N}$, stops when these elements are at equilibrium. The energy production rate is proportional to $T^{19}$, destabilizing the medium and creating a convective zone. }}. The normalized gravity profile, changes from a parabolic profile for the fully convective model to a hyperbolic one in the upper half of the stellar bulk once the radiative core appears and the matter gets more and more concentrated toward the centre. This evolution of the model's structure
is also visible in the middle panel of Fig. \ref{fig:kippenhahn}, which shows the convective envelope's aspect ratio $\xi=r_{\rm i}/r_{\rm o}$. Here, the inner radius $r_{\rm i}$ is the radius at the bottom of the envelope while outer radius $r_{\rm o}$ is taken as $90\%$ of the total radius. This choice allows to focus on the adiabatic part of the envelope and neglect dramatic drops of density close to the surface\footnote{Large scale 3D numerical simulations as performed in Sect.~\ref{section:dipole_stability} generally neglect these surface layers for numerical sobriety.}. \anna{We see that the thickness of the radiative zone increases during the WTTS phase and so $\xi$ increases until it reaches a plateau that will last during all the main sequence (MS).} 
The density contrast $N_\rho \equiv \log [\rho(r_{\rm i})/\rho(r_{\rm o})]$ in the convective envelope follows the inverse behaviour: it decreases as most of the matter gets concentrated in the central region. This counterintuitive trend results from the simultaneous outward migration of the base of the convective zone (i.e., the increase in $\xi$). For the very-low-mass star considered here, it goes from $4$ to about $2.5$. We note that for a solar mass, $N_\rho$ gets slightly lower values around $1.5$.

The characteristic timescale of the PMS star is of the order of the global thermal (or Kelvin-Helmholtz) timescale $\tau_{\rm KH} \equiv GM^2 /(RL)$, where $G$ is the gravitation constant, and $M$, $R$ and $L$ are the stellar mass, radius and luminosity. This is the time necessary to radiate away the gravitational energy released during contraction. This timescale is always orders of magnitude larger than the period of rotation $P_{\rm rot}$ or the convective turnover timescale $\tau_{\rm conv}$, which is provided by the mixing length theory (MLT) and integrated over the convective bulk. As shown by observations and simulations, the timescale associated with stellar convective dynamo is of the order of thousands of rotation periods only, so that we can safely neglect the variations of the radius and aspect ratio in these simulations. Moreover, in this figure, we also see that the convective Rossby number averaged over the convective zone, representing the ratio $P_{\rm rot}/\tau_{\rm conv}$, tends to increase globally along the evolution on the PMS, going from about $0.01$ to about $0.1$: it is thus a good indicator of age on the PMS. We note however that this monotonic behaviour can be slightly perturb depending on the assumptions considered for the angular momentum redistribution \citep[e.g.,][]{Amard2023}.

Finally, the very small predicted values of $\tau_{\rm conv}$ are generally used as a justification for solid body rotation in convective zones, as assumed by \cesamxx. This is, however, a rough simplification. In contrast physically-grounded models (e.g., based on conservation of specific angular momentum) allow for decreasing rotation rate with radius in thick envelopes \citep[e.g.,][]{Kissin2015}. \anna{Contrary to \cesamxx assumption of a solid body rotating convective zone}, radial differential rotation is certainly ubiquitous in PMS convective envelopes, and mainly ruled by the rotation of the underlying radiative layers. This latter evolves on a $\tau_{\rm KH}$ timescale and thus can be considered as constant with respect to convective dynamos. Our aim is thus to study how this differential rotation affects dynamo in the convective layers using direct numerical simulations (DNS).

\section{Dipole stability in sheared 3D DNS}
\label{section:dipole_stability}

In this section, we investigate the influence of an imposed radial differential rotation on dipolar dynamos driven by convection using DNS.
\subsection{Numerical setup  and parameters choice}
We solve the so-called anelastic dynamo model in MagIC software \citep{jones2011anelastic,gastine2012effects}, solving coupled equations of momentum, magnetic induction and entropy in spherical shells rotating with a surface rotation rate $\Omega_s$.
Convection is driven by the gradient of entropy imposed between the inner and outer spheres as Dirichlet boundary conditions. These surfaces are considered insulating to provide boundary conditions for magnetic field. 

At the top and bottom boundaries of the shells, we assume no-slip conditions. \anna{In the frame co-rotating with the surface at angular velocity $\Omega_s$, we introduce a difference in rotation rate imposed between the inner and the outer spherical shell, $\Delta \Omega$, modelling the combined effects of the contraction-induced speed-up of the inner core, the surface angular momentum loss by magnetic interactions with the surroundings, and the transport of the angular momentum inside the star. This way, the angular velocity below the convective zone is $\Omega_s+\Delta \Omega$, with $\Delta \Omega>0$}. \anna{Even without imposed external differential rotation, i.e. $\Delta \Omega = 0$, weaker differential rotation naturally develops in convective dynamos as a result of momentum transport by convective turbulent motions and magnetic stresses. Its development is a nonlinear process, and is not taking into account external to convection processes (such as differential rotation of the radiative core and winds). Imposing differential rotation across the convection zone allows us to model and control the impact of these processes on the stellar dynamo.} 

The two key governing parameters of the system are shear Rossby number,
\begin{equation}\label{eq:rosh}
 \Ross_{\rm sh} = \Delta \Omega /\Omega_s,
\end{equation}
and (local) 
convective Rossby number, defined as
\begin{equation}\label{eq:roc}
    \Ross_{\rm conv} (r) = \frac{u_{\rm rms} (r)}{l_{\rm conv} (r) \Omega_s}, \quad l_{\rm conv} (r) = \frac{ \pi E^{\rm kin} (r)}{\sum_\ell \ell E^{\rm kin}_{\ell} (r)},
\end{equation}
based on convective length scale $l_{conv}(r)$; $u_{\rm rms}$ is the rms velocity of the flow, and $E^{\rm kin}_\ell(r)$ is the contribution of the spherical harmonic degree $\ell$ to the total kinetic energy $E^{\rm kin}(r)$. If we average this quantity over radius, we get $Ro_{\rm conv}^\ell$, as defined in \cite{christensen2006scaling}, which depends on the radially-averaged characteristic length scale and rms velocity of the flow. 
While shear Rossby number is the control parameters of our setup, the convective Rossby number is the diagnostic parameter and is not known \textit{a priori}, before the simulation has been performed. 

\begin{figure}[t]
\centering
         \includegraphics[width=0.41\textwidth]{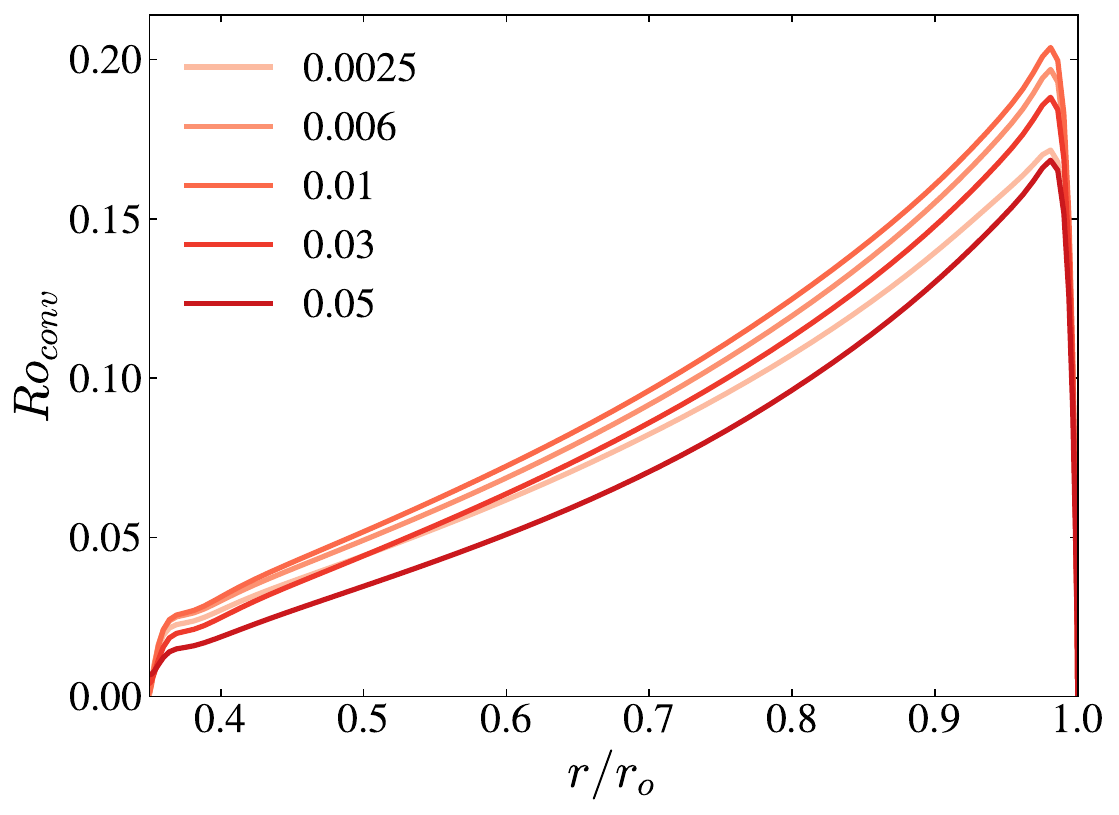}
\caption{Local convective Rossby number as a function of normalized radius $r/r_{\rm o}$ for typical values of $\Ross_{\rm sh}$ considered in this work. The gravity profile is taken from \cesamxx, approximated by Eq.~\eqref{eq:g_r_cesam}. See run \texttt{gc\_1} in Table~\ref{tab:sim_param} for more details on the parameters.}
    \label{fig:Roconv_varom}
\end{figure}

 The Ekman number is set to $\Ekma = 10^{-4}$, corresponding to moderately rotating stars, and Prandtl number is set to $\Pran=1$. Although the latter quantities are much larger than values expected in stars, yet it is likely a valid choice if we consider the diffusivities of the simulations as effective turbulent coefficients \citep[e.g.,][]{bhattacharya2025scale}. We restrict our simulations to layers between the bottom of the convective envelope at radius $r_{\rm i}$ and the external radius $r_{\rm o}$ located $10$\% of the actual stellar radius below the surface. Without loss of generality, the reference state of the simulations follow a polytropic structure (with a polytropic index $n=2$). The density contrast is set to $N_\rho = 2.5$, equivalent to an inner-to-outer density ratio of $\rho(r_{\rm i})/\rho(r_{\rm o}) \approx 12$ typical of PMS stars (see Sect.~\ref{section:early_stell_evol}).
 Furthermore, we vary radius (aspect) ratio $\xi = r_{\rm i}/r_{\rm o} \in [0.1, 0.2, 0.35]$ to mimic the narrowing of the convective envelope during the PMS evolution. Magnetic Prandtl number is $\Pram \in [4,6]$, and we employ several values of the Rayleigh number $\Rayl$ to test more or less turbulent convective states. \anna{See Appendix~\ref{sec:appA} for the definitions of these parameters.}

In addition to the evolutionary constraints on the density contrast and the aspect ratio, we also take into account changes in the gravity profile during the PMS evolution. As an illustration, Fig.~\ref{fig:cesam}a shows the normalized gravity profiles of a $0.8 M_\odot$ star at solar metallicity and at two different stages (see also Fig.~\ref{fig:grav_omega} for a $0.55M_\odot$ star). The gravity profile evolves from a quasi parabolic (e.g., 1 Myr T Tauri model with $N_\rho \simeq 1.7$) to a predominantly hyperbolic one $g\propto 1/r^2$ (e.g., at the age of $4.9\times10^3$ Myrs on the early MS, with $N_\rho \simeq 1.5$). Regarding the very beginning of stellar formation, it is also interesting to consider the limiting case of a uniform density spherical shell, with the gravity increasing linearly with radius. Therefore, for sake of completeness, we explore three different profiles in our simulations: $g\propto r$, $g\propto 1/r^2$ and a parabolic-like gravity profile typical of young PMS stars (i.e., around 1 Myr). This latter profile is taken from the \cesamxx model in Fig.~\ref{fig:cesam}a and approximated with polynomials as a function of $\mu \equiv r/r_{\rm o}$,
\begin{equation}\label{eq:g_r_cesam}
    g(r) \approx -0.05  + 4.05 \mu -3.27 \mu^2 - 2.39 \mu^3 + 2.18 \mu^4.
\end{equation}

The initial conditions of our simulations correspond to dipolar dynamos already identified in \citet{GastineDW12}, who already considered the effects of $\Rayl$ and $N_\rho$ for two different aspect ratios on the stability of dipolar dynamos. Our parameter set is very similar except that we consider higher values of $\Pram$ to stabilize dipolar dynamos, as required by our higher (and more realistic) value of $N_\rho$. We then impose a radial shear rate $\Delta \Omega$. Previous work show that differential rotation can give rise to different dynamical regimes in stably-stratified spherical Couette flows under the effect of shear instability~\citep{wicht2014flow}. Here, we limit our attention to low values of shear rates below the critical one for the shear instability; convection is thus the primary instability responsible for turbulent motions which are not considerably perturbed by the imposed shear, as can be seen from relatively unchanged radial distribution of $\Ross_{\rm conv}$ in Fig.~\ref{fig:Roconv_varom}.

\begin{figure*}[t]
\centering
     \begin{subfigure}[b]{0.48\textwidth}
         \centering
         \includegraphics[width=\textwidth]{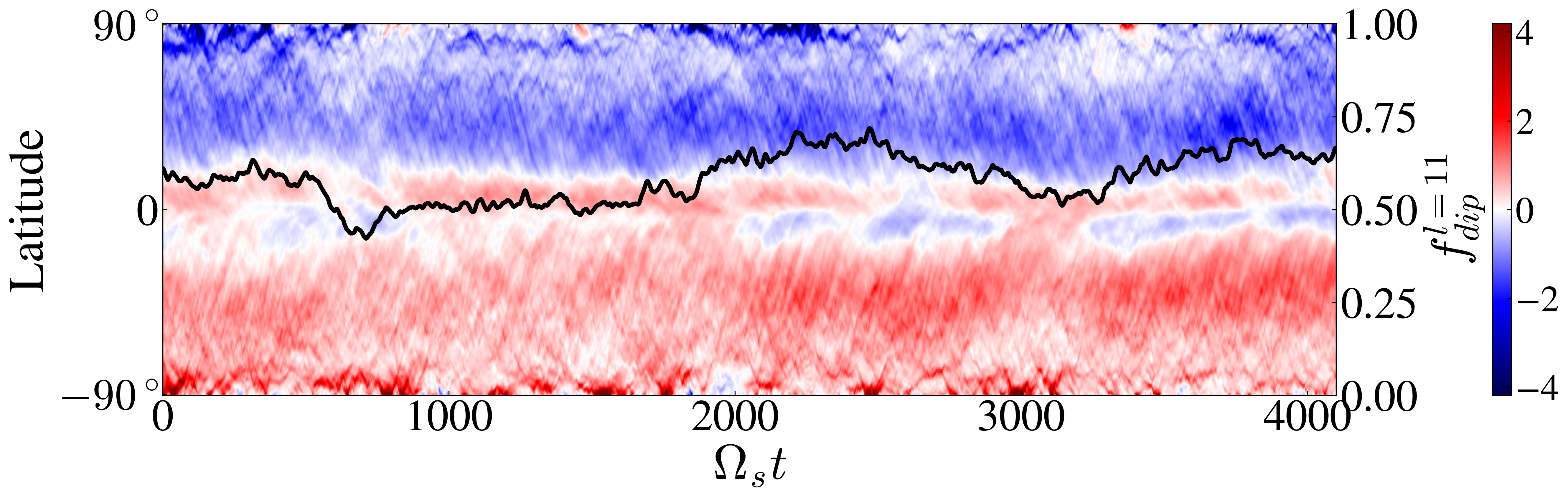}
         \caption{}
     \end{subfigure}
     \hfill
          \begin{subfigure}[b]{0.48\textwidth}
         \centering
         \includegraphics[width=\textwidth]{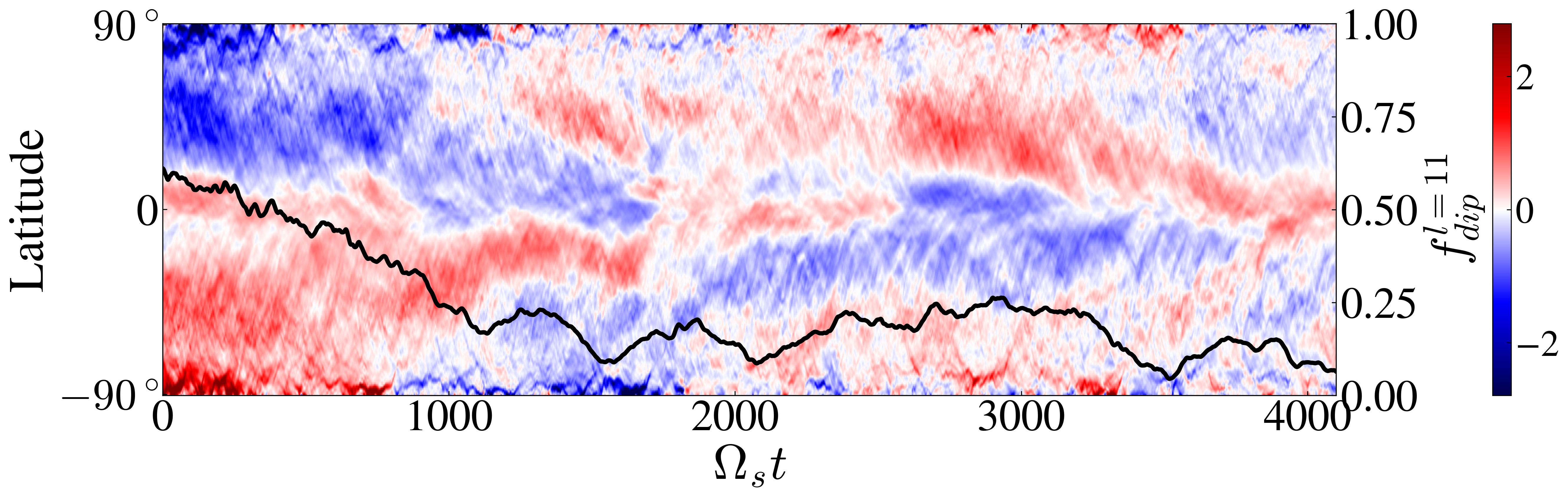}
         \caption{}
     \end{subfigure}
     \vfill

          \begin{subfigure}[b]{0.48\textwidth}
         \centering
         \includegraphics[width=\textwidth]{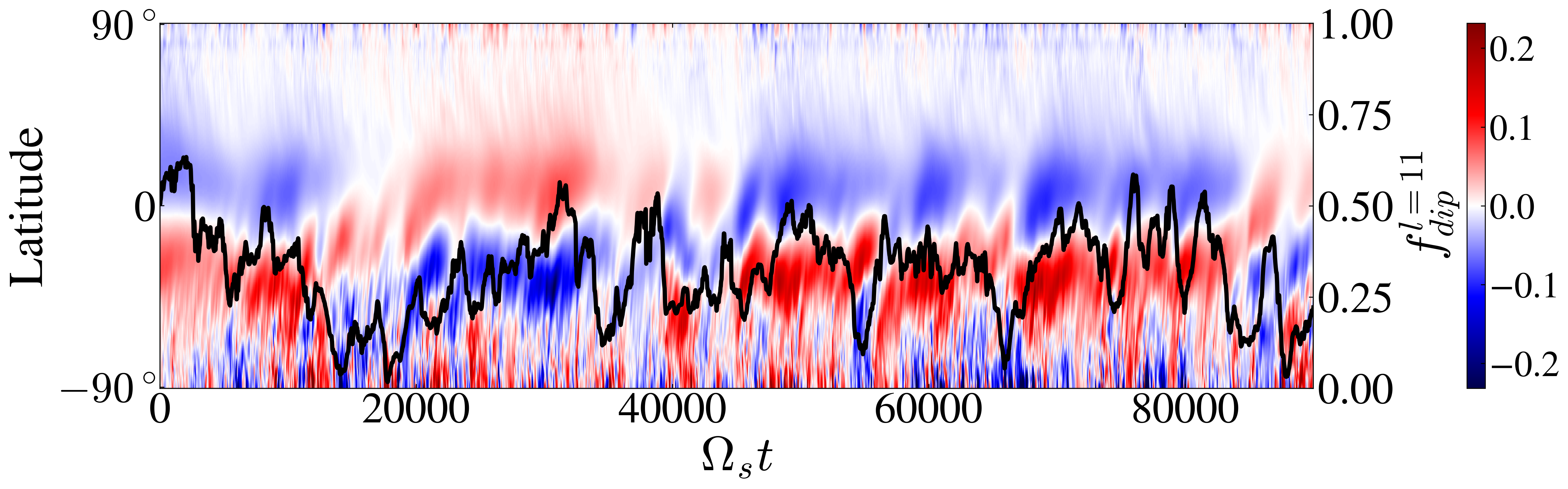}
         \caption{}
     \end{subfigure}
     \hfill
          \begin{subfigure}[b]{0.48\textwidth}
         \centering
         \includegraphics[width=\textwidth]{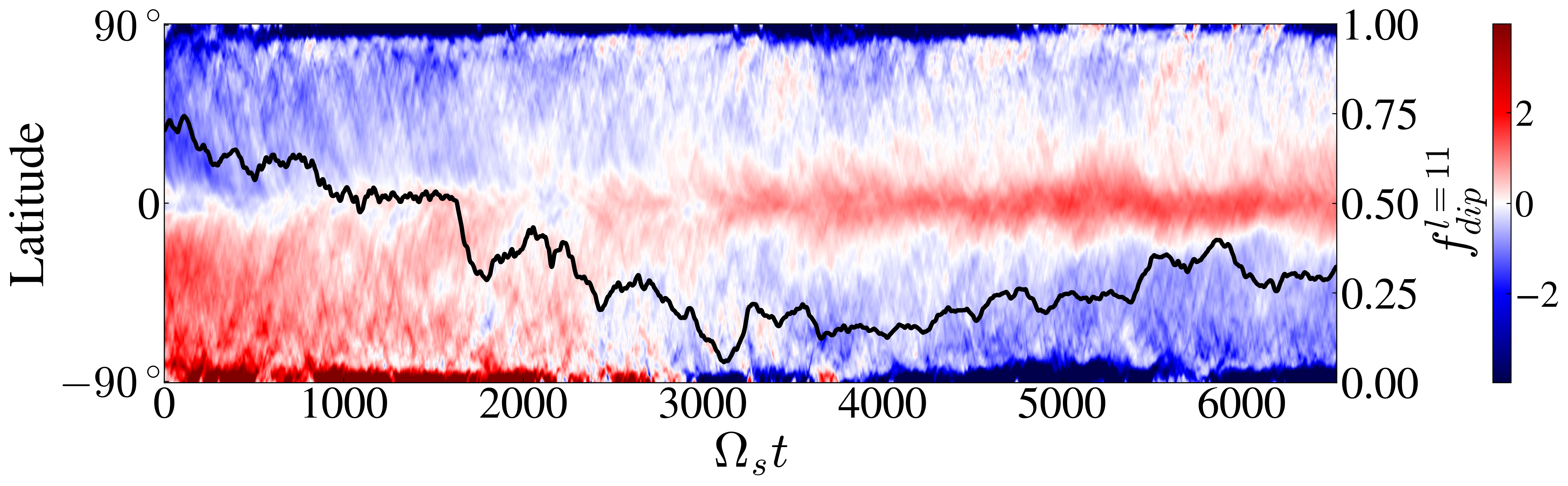}
         \caption{}
     \end{subfigure}
\caption{Butterfly diagrams of the axisymmetric radial magnetic field component, $\langle B_r \rangle_\phi$, together with the dipolarity. \anna{$\Omega_s t$ corresponds to the rotation time, and follows the uniform rotation of a meridional plane.} (a) Stable dipole for $g\propto 1/r^2$, $\Rayl=5\times10^6$, $\xi = 0.35$, $\Ross_{\rm sh} =0.005$. (b) Equatorially propagating dipolar waves for the same parameters, except for  $\Ross_{\rm sh} =0.02$. (c) Aperiodic dynamos for $g\propto 1/r^2$, $\Rayl=5\times10^5$, $\xi = 0.1$, $\Ross_{\rm sh} = 0.015$. (d) Steady quadrupolar dynamos for $g\propto r $, $\Rayl=1.6\times10^7$, $\xi = 0.2$, $\Ross_{\rm sh} = 0.3$. \anna{Magnetic field was normalized with $(\rho_o \mu_0 \lambda_i \Omega_s^{1/2})$, where $\rho_o$ is the density at the outer boundary, $\lambda_i$ is magnetic diffusivity at the inner boundary, and $\mu_0$ is  the free space permeability.}}
    \label{fig:fdip_time}
\end{figure*}

\subsection{Patterns of convection with different gravity profiles}
\input{32_charly}

\subsection{Evolution of dipolarity and magnetic topology}

Figure~\ref{fig:fdip_time} shows the evolution of the axisymmetric radial component of magnetic field, $\langle B_r \rangle_\phi$, as a function of time and latitude for the most typical behaviours observed in our simulations. In the following, we quantify  the dipolarity  using the integral value,
\begin{equation}\label{eq:fdip}
    f_{\rm dip}^{\ell_{\rm cut}} =\left(\dfrac{\int_{r_{\rm i}}^{r_{\rm o}} \sum_{m=-1}^{m=1} E^{\ell=1,m}_{\rm mag} (r) r^2 dr}{ \int_{r_{\rm i}}^{r_{\rm o}}\sum_{\ell=1}^{\ell_{\rm cut}} \sum_{m=-\ell}^{m=+\ell} E^{\ell,m}_{\rm mag} (r) r^2 dr}\right)^{1/2} \; ,
\end{equation}
comparing the energy concentrated in the $\ell=1$ dipolar modes and the energy of large-scale spherical harmonics up to a cut-off angular degree ($\ell_{\rm cut}=11$ in this work).

When $\Ross_{\rm sh}$ is weak, we conserve a stable dipole magnetic topology which does not vary in time, as seen in Fig.~\ref{fig:fdip_time}a for $\Ross_{\rm sh}=0.005$. We note that the dipolarity slightly decreases in comparison to non-perturbed case (i.e., $\Ross_{\rm sh}=0$) yet remains higher than $0.5$. For the same gravity profile, $g\sim 1/r^2$ and slightly higher shear, i.e. $\Ross_{\rm sh} = 0.02$, the dipole become distorted. After the transition, dipolarity remains lower than $0.5$ as the flow typically exhibits equatorward propagating waves (see Fig.~\ref{fig:fdip_time}b)
We also sometimes observe sharper and erratic dipole-quadrupole transitions, as  seen in Figs.~\ref{fig:fdip_time}c. Further increase in $\Ross_{\rm sh}$ usually does not change this distribution of magnetic field. When $\Ross_{\rm sh}$ is very large and rotational dynamics dominates, we can nevertheless observe steady dipolar-quadrupolar solutions with dipole mostly concentrated around the
tangent cylinder. Such solutions would be relevant only for very rapidly rotating stars, not low-mass PMS stars; they are therefore not represented here and not considered in the following.


\subsection{Dipole collapse: shear versus convection competition}
\label{sect:compet}

\begin{figure}
    \centering
    \includegraphics[width=0.98\linewidth]{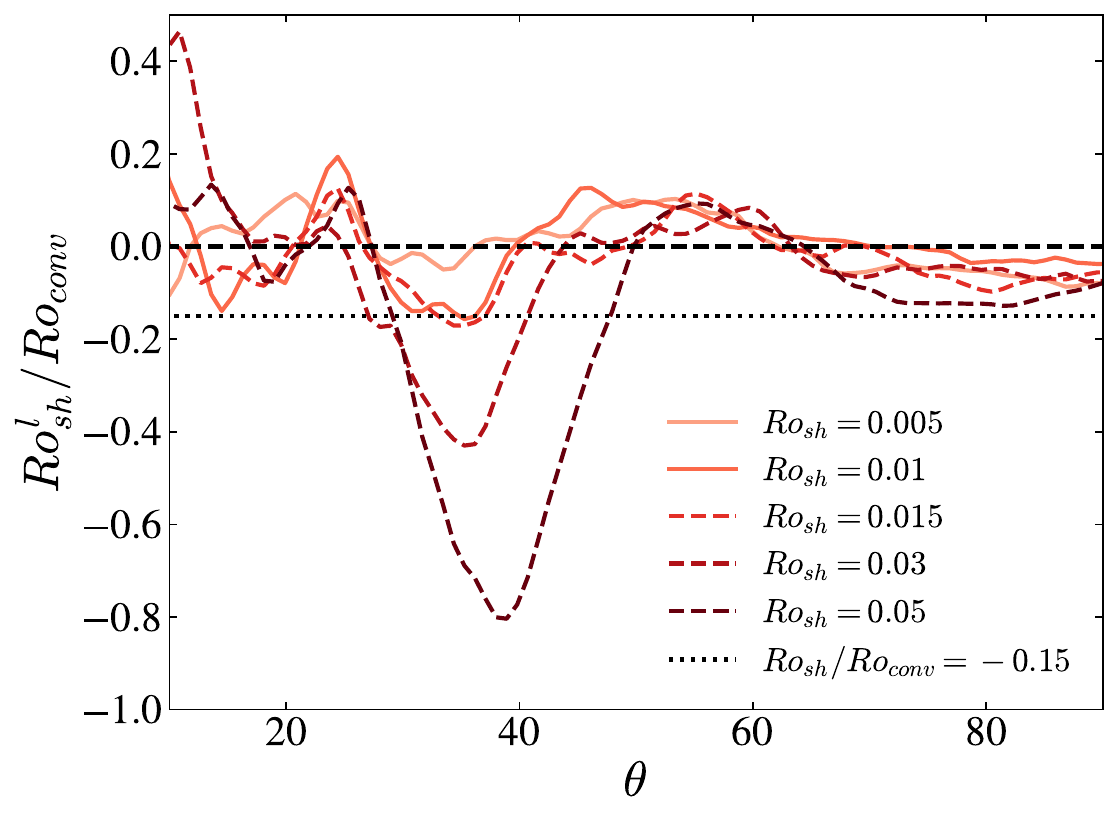}
    \caption{Ratio of the local radial shear to the local convective Rossby number as a function of latitude (in degree) at the midgap, $r/r_o = 0.5$, for varying $\Ross_{\rm sh}$. Solid lines denote stable dipolar states as those in Fig.~\ref{fig:fdip_time}a, and dashed lines denote collapsed dipoles as those in Fig.~\ref{fig:fdip_time}b.}
    \label{fig:Rosh_Roconv_theta}
\end{figure}  

The angular velocity distribution of our simulations with $\Ross_{\rm sh}=0$ corresponds to a configuration with slow poles and fast equator or near-equator jets \citep[e.g.,][]{Gastine2012zonal}. Thus, the local radial gradient of differential rotation, or local shear Rossby number,
\begin{equation}\label{eq:rosh_rad_2d}
    \Ross_{\rm sh}^l \sim \frac{r}{\Omega} \diff{\Omega}{r},
\end{equation}
typically increases with radius, with most pronounced increase at the tangent cylinder circumscribing the inner sphere (Fig.~\ref{fig:Rosh_Roconv_local}a). \anna{Here, $ \Omega = \langle U_\phi \rangle /r \sin \theta$, where $U_\phi$ the azimuthal component of velocity in the frame co-rotating with angular velocity $\Omega_s$ of the surface and $\langle \cdots \rangle$ denotes the azimuthal average. $ \Omega$ satisfies the boundary conditions in our simulations, i.e. $ \Omega (r_i) = \Delta \Omega$, and $\Omega (r_0) = 0$.} With non-null imposed rotation $\Delta \Omega$, all the area around the inner sphere rotates faster than background rotation $\Omega_s$, with developing so-called Stewartson  shear layer~\citep{wicht2014flow}. This leads to suppression of the area of positive differential rotation at this location, and development of negative rotation gradients in the radial direction instead (Fig.~\ref{fig:Rosh_Roconv_local}b). As $\Ross_{\rm sh}$ increases, the shear  becomes more and more pronounced and extended in this region until the dipolar dynamo is disrupted (Figs.~\ref{fig:Rosh_Roconv_local}c and d). This process can be seen as a competition between local shear expressed by $\Ross_{\rm sh}^l(r)$, and convection, expressed by (local) convective Rossby number $\Ross_{\rm conv} (r)$, which vary with radius (e.g., Fig.~\ref{fig:cesam}b). Figure~\ref{fig:Rosh_Roconv_theta} shows the ratio of these two quantities for $g\sim1/r^2$ and $\xi=0.35$, corresponding to narrow convective zone, for several values of global $\Ross_{\rm sh}$, in the middle of the domain, $r_{\rm i}/r_{\rm o}=0.5$ and as a function of latitude $\theta$. As discussed above, we see a clear correlation between the strength of the negative shear, developing at $\theta \in [30-40]$, and the collapse of the dipole (dashed lines), which takes place at $\Ross_{\rm sh} \approx 0.15$ in this case. This result supports the idea that this ratio represents the appropriate parameter to characterize the dipole collapse.

\begin{figure*}[t]
     \centering
               \begin{subfigure}[b]{0.46\textwidth}
         \centering
         \includegraphics[width=\textwidth]{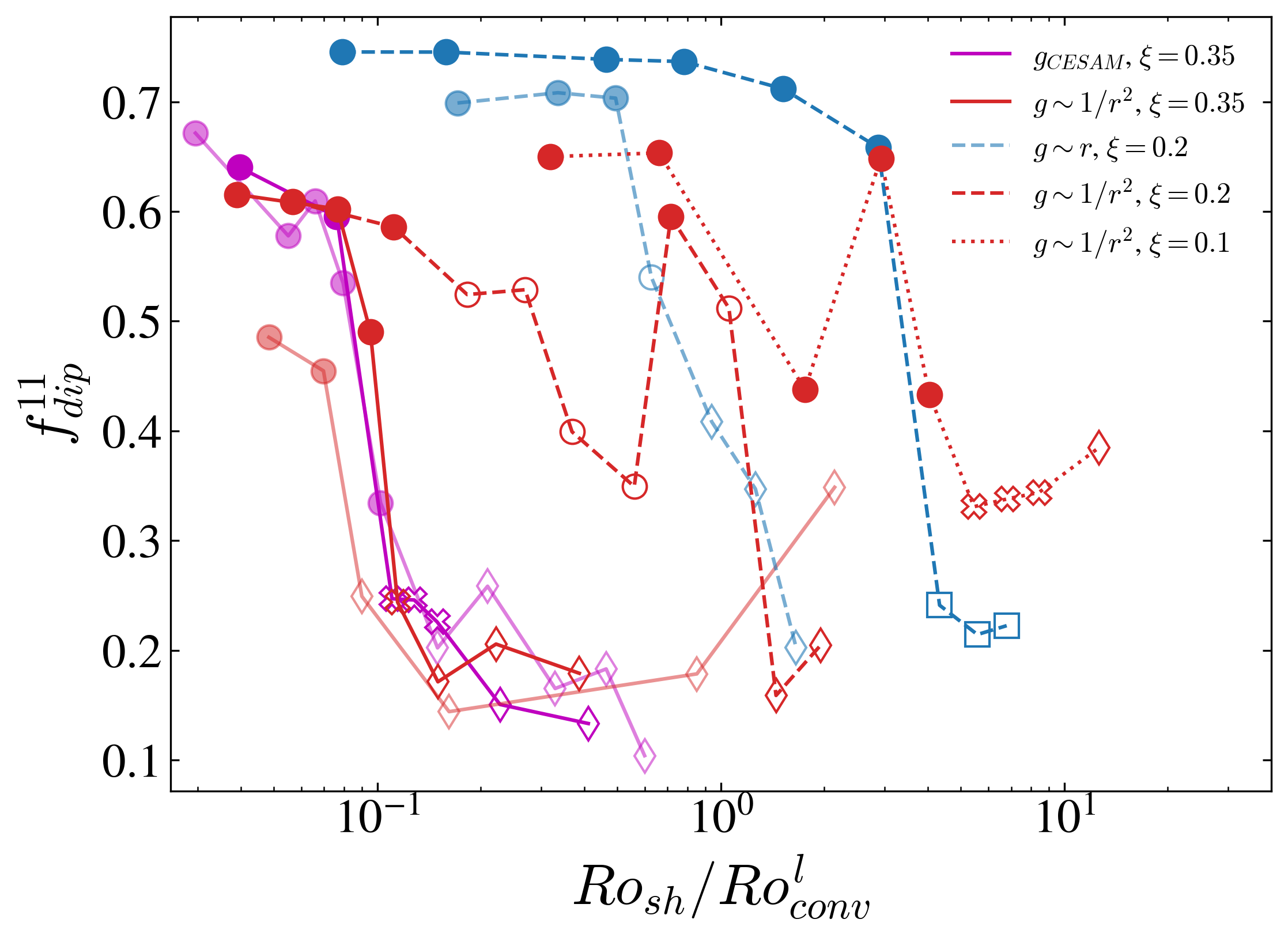}
         \caption{}
     \end{subfigure}
     \hfill
          \begin{subfigure}[b]{0.46\textwidth}
         \centering
         \includegraphics[width=\textwidth]{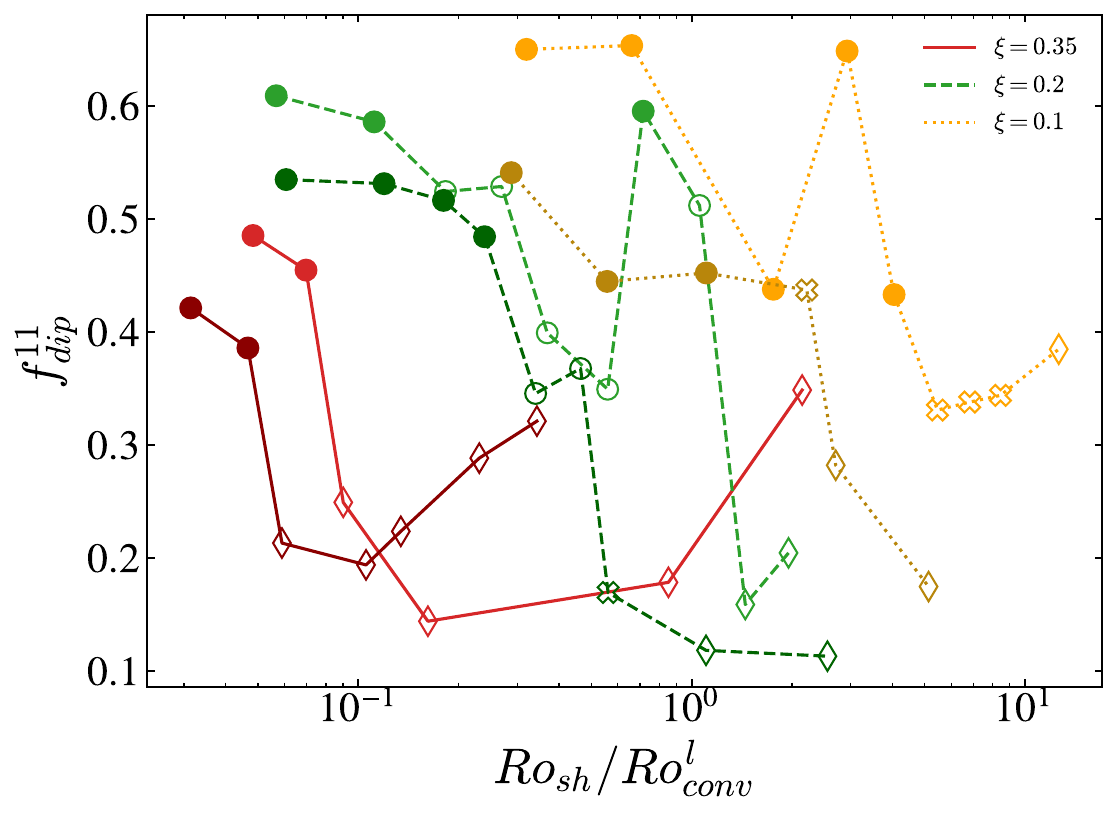}
         \caption{}
     \end{subfigure}
     \vfill
             \begin{subfigure}[b]{0.46\textwidth}
         \centering
         \includegraphics[width=\textwidth]{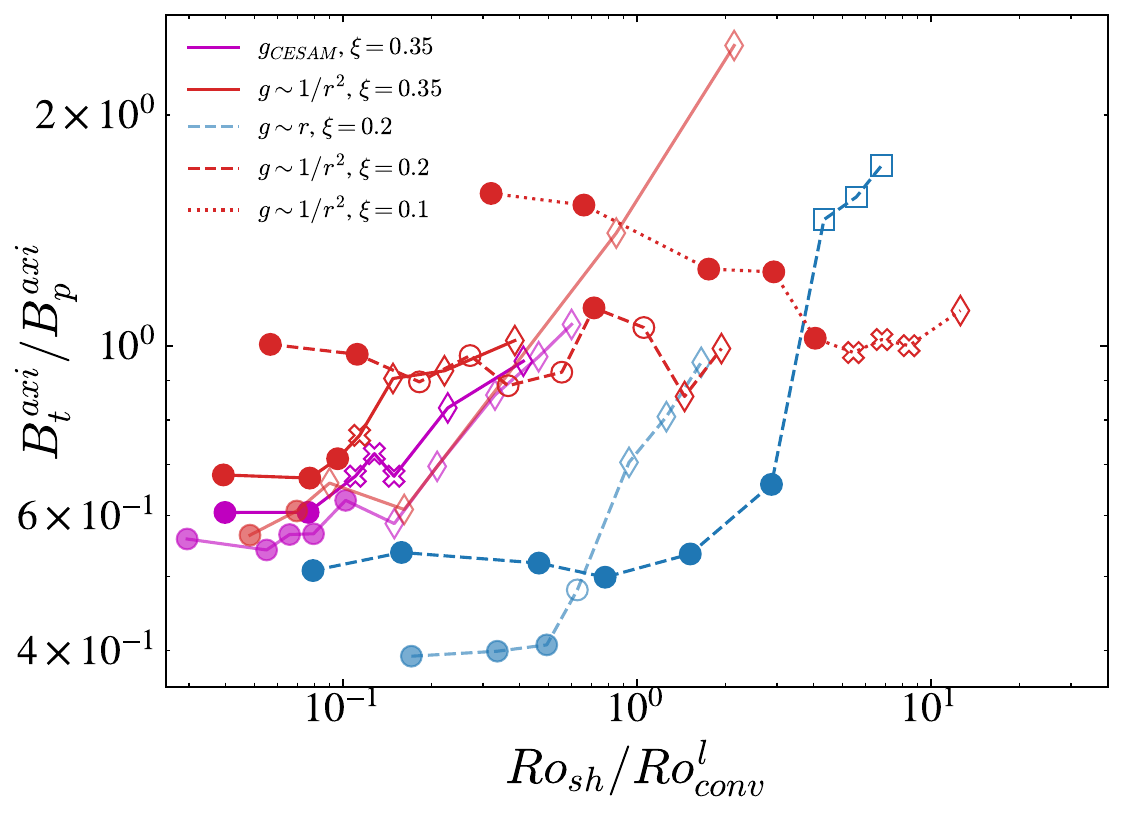}
         \caption{}
     \end{subfigure}
     \hfill
    \begin{subfigure}[b]{0.45\textwidth}
         \centering
         \includegraphics[width=\textwidth]{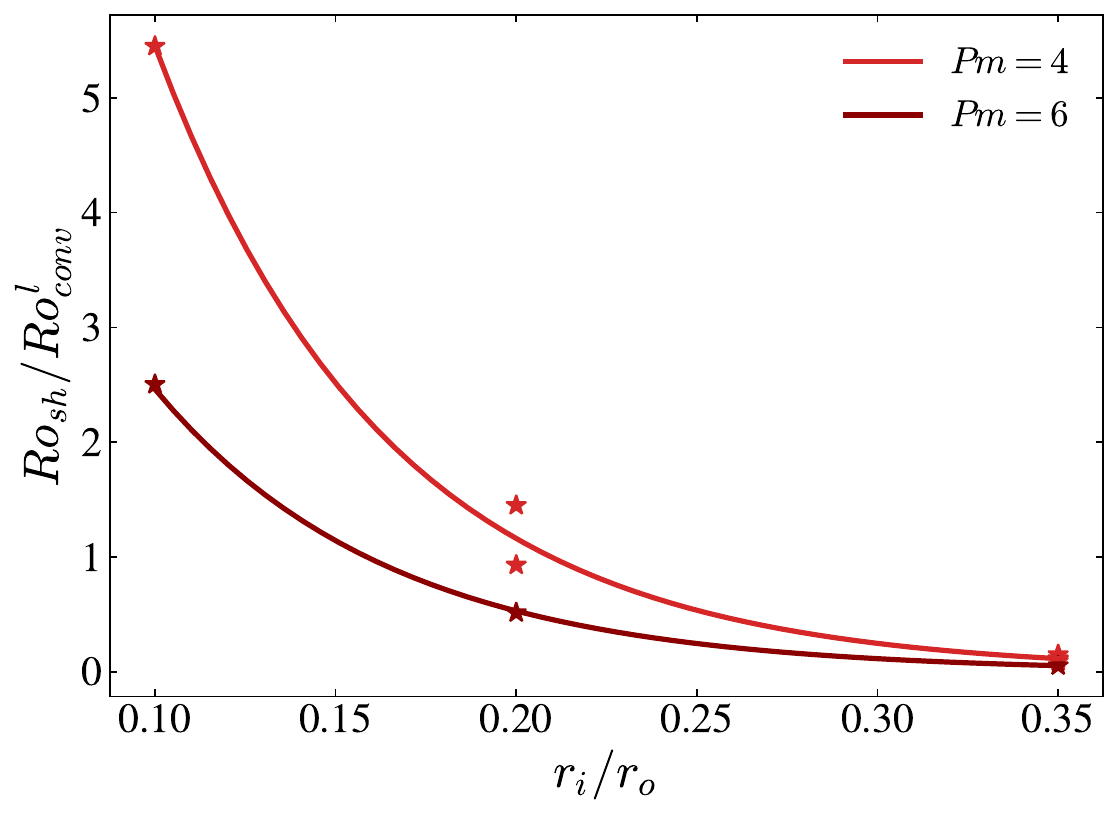}
         \caption{}
     \end{subfigure}
     \caption{(a) Time-averaged dipolarity parameter as a function of the ratio between shear and convective Rossby numbers, for $\Pram=4$ and three different gravity profiles: $g(r)$ from CESAM in Eq.~\eqref{eq:g_r_cesam} in magenta, $g\propto 1/r^2$ in red, and $g\propto r$ in blue. Solid lines, $\xi = 0.35$, dashed, $\xi =0.2$, dotted, $\xi = 0.1$. More (less) transparent lines corresponds to the runs with lower (higher) $\Rayl$ number, for fixed gravity profile and radius ratio (see Table~\ref{tab:sim_param}). Filled circles: stable dipoles (Fig.~\ref{fig:fdip_time}a), empty circles: reversing dipoles, diamonds: dipolar-quadrupolar waves (Fig.~\ref{fig:fdip_time}b), crosses: mixed dipolar/multipolar solutions without coherent temporal dynamics (Fig.~\ref{fig:fdip_time}c), squares: equatorially symmetric solutions (Fig.~\ref{fig:fdip_time}d).  (b) Same as in panel (a) for $g\propto 1/r^2$ only and varying radius ratio. Light color, $\Pram=4$; dark color $\Pram=6$. Line styles and symbols as in panel (a). (c) Ratio of rms of axisymmetric toroidal and poloidal magnetic components, $B^{\rm axi}_t$ and $B^{\rm axi}_t$. Runs, lines and symbols from panel (a). (d) Estimated critical collapse value of $\Ross_{\rm sh}/\Ross_{\rm conv}^\ell$ from all curves in panels (a) and (b), as a function of $\Pram$ and radius ratio $\xi$ (stars), and the corresponding exponential fits (lines, Eq.~\ref{eq:gu_fit_form}). } 
    \label{fig:ro_sh_ro_conv}
\end{figure*}

 To simplify and for future applications in one-dimensional stellar evolution codes (see example in Sect.~\ref{section:discussion}), we seek instead for a criterion of the dipole collapse as a function of the total shear Rossby number, $\Ross_{\rm sh}$, as defined by Eq.~\eqref{eq:rosh}, and the integrated in radius convective Rossby number  $\Ross_{\rm conv}^\ell$, obtained from the radial average of Eq.~\eqref{eq:roc}. Fig~\ref{fig:ro_sh_ro_conv}a shows the time-averaged dipolarity of our models as a function of $\Ross_{\rm sh}/\Ross_{\rm conv}^\ell$,  as soon as the dynamo steady state is reached. We first see that for a fixed aspect ratio $\xi=0.35$ and for both the hyperbolic gravity profile and the realistic gravity profile from \cesamxx, the transition from steady dipole toward dipolar oscillatory dynamos occurs at $\Ross_{\rm sh}/\Ross_{\rm conv} \sim 0.1$, which is very close to the local value observed in Fig.~\ref{fig:Rosh_Roconv_theta}. In addition, this threshold does not evolve when varying the $\Rayl$ number. Figure~\ref{fig:ro_sh_ro_conv}b compares solutions for different envelope thickness (i.e., $\xi=0.1$, $0.2$ and $0.35$). The thicker the envelope, the higher the critical value of $\Ross_{\rm sh}/\Ross_{\rm conv}$. Moreover, the transition seems to occur less abruptly at $\xi=0.2$, that is, over a wider interval of values for $\Ross_{\rm sh}/\Ross_{\rm conv}$. In this interval, we observe reversing yet dipolar solutions with a seeming bi-stability with the steady dipole. For $\xi=0.1$, the dipole collapse is even more delayed, with the flow first passing through a range of distorted states with slow evolution and no clear frequency, as those shown in Fig.~\ref{fig:fdip_time}c. These solutions eventually develop a periodic frequency and become oscillating waves for higher value of $\Ross_{\rm sh}$.

A particular case is the homogeneous mass distribution profile with  $g\propto r$ in Fig.~\ref{fig:ro_sh_ro_conv}a (blue color). This case shows signs of bi-stability, with low values of $\Rayl$ following the 
same trend observed with other gravity profiles for thin envelopes (i.e., $\xi = 0.35$),
and high values of $\Rayl$ leading to a very stable dipolar branch that is lost only for very strong values of $\Ross_{\rm sh}$ toward a non-oscillatory, equatorially symmetric (quadrupolar) dynamo branch (Fig.~\ref{fig:fdip_time}d). 
We will come back to this phenomena in more details when discussing the mechanisms of dynamo collapse in the next section.

\subsection{Increasing magnetic effects and criterion for dipole stability }
Previous works highlighted that increasing magnetic Prandtl number tends to promote strong-field dipolar solutions and therefore stabilizes the dipole against multipolar, disordered solutions \citep{schrinner2014topology,dormy2016strong,petitdemange2018systematic,menu2020magnetic,pinccon2024coriolis}. In this work, we tested this effect to identify if it will  increase the critical $\Ross_{\rm sh}/\Ross_{\rm conv}^\ell$. Counterintuitively, we find that increasing $\Pram$ leads to slightly earlier destabilization of the dipole, for all tested radii (Fig.~\ref{fig:ro_sh_ro_conv}b), and at comparable values of $\Rayl$. Since in DNS $\Pram$ controls the strength of magnetic field, this result indicates that stars with stronger field (as, for example, measured by their Elsasser number), are more prone to lose their dipoles.  A possible explanation for this behavior may result from a better coupling between the super-rotating inner core with the bulk flow. In other words, magnetic effects are likely to extend the influence of the differential rotation on a larger distance. We leave detailed investigation of this effect for the future work.

\begin{figure*}[h!]
     \centering
       \begin{subfigure}[b]{0.2\textwidth}
         \centering
        $B_r$
         \includegraphics[width=\textwidth]{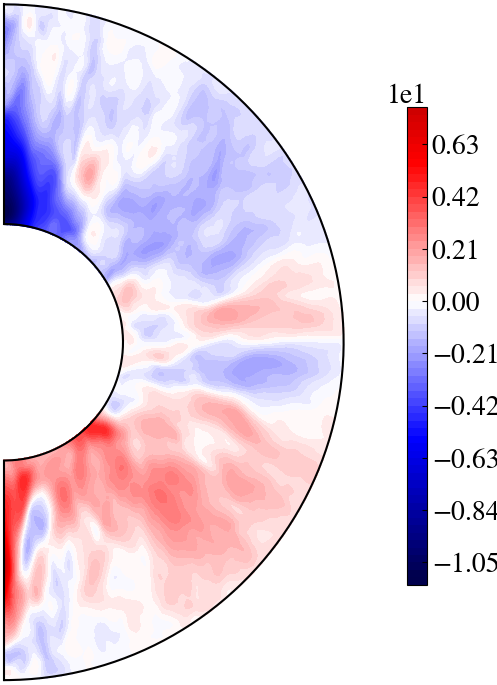}
         \caption{}
     \end{subfigure}
               \hfill   
        \begin{subfigure}[b]{0.2\textwidth}
         \centering
                  Helicity
         \includegraphics[width=\textwidth]{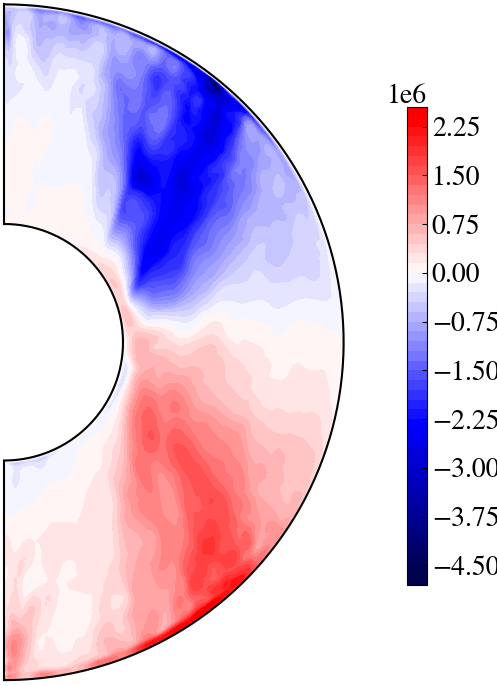}
         \caption{}
     \end{subfigure}
        \hfill
    \begin{subfigure}[b]{0.2\textwidth}
         \centering
           $\gamma_r$
         \includegraphics[width=\textwidth]{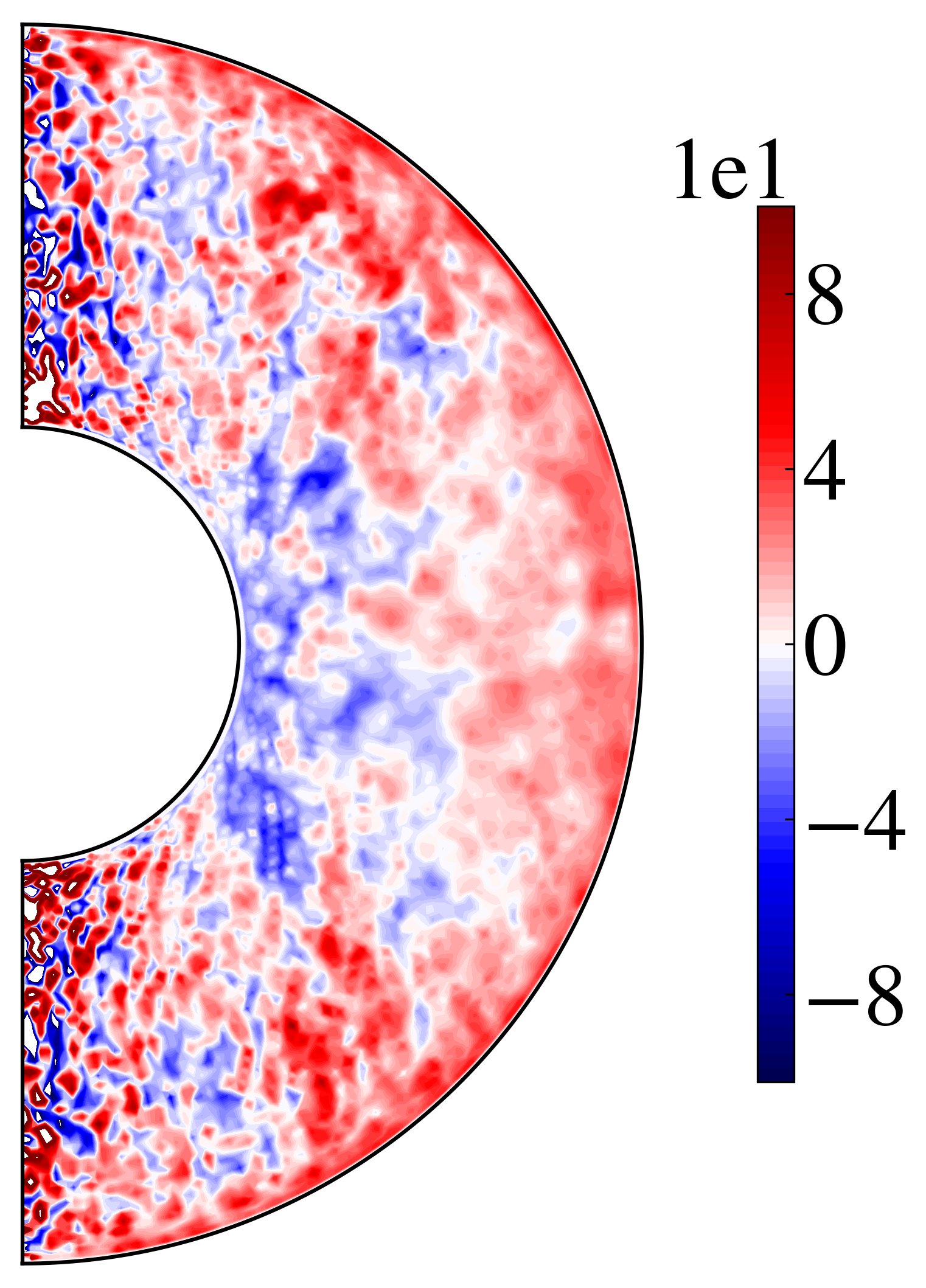}
         \caption{}
     \end{subfigure}
               \hfill        
               \begin{subfigure}[b]{0.22\textwidth}
         \centering
          $\langle U_r\rangle_\phi$
         \includegraphics[width=0.94\textwidth]{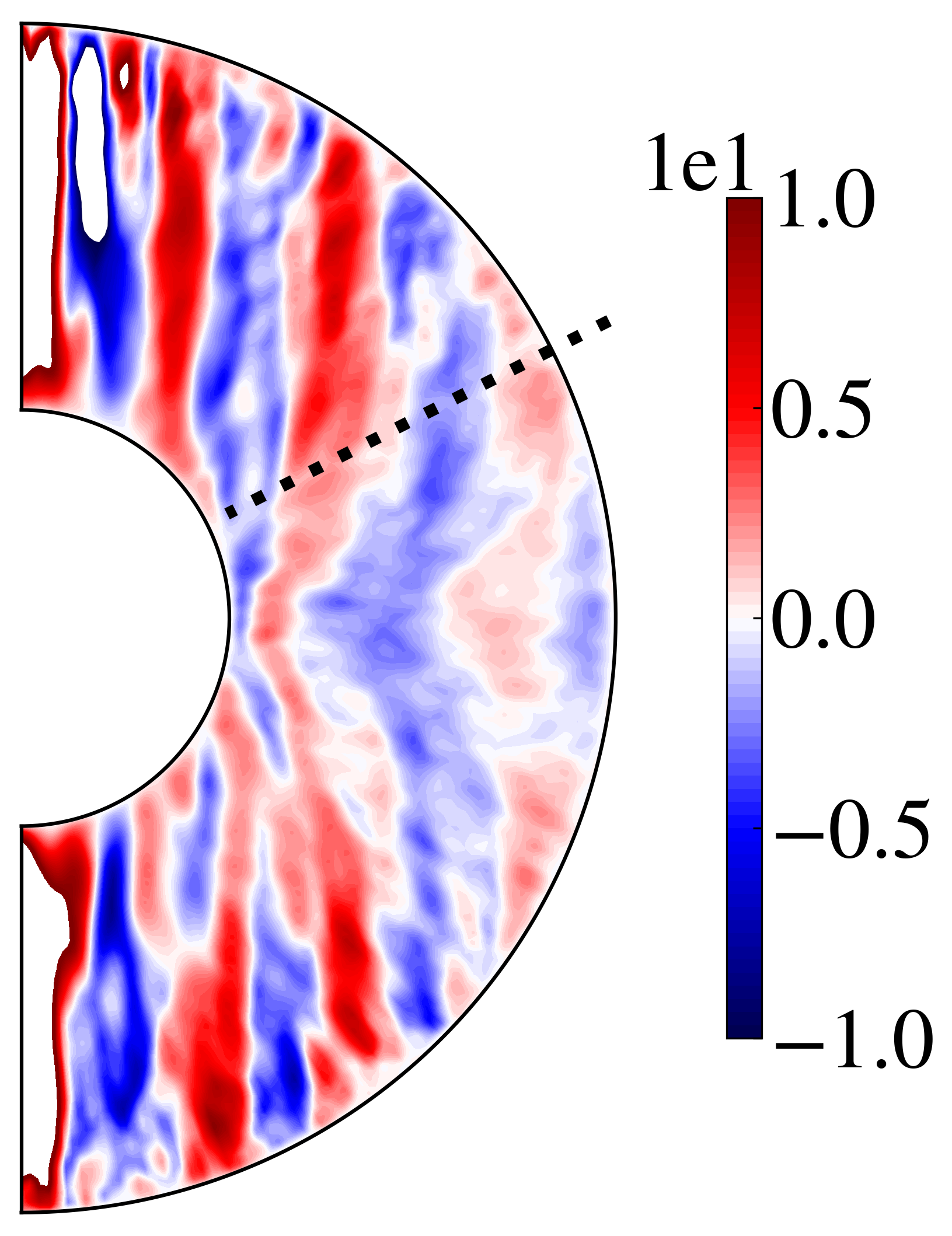}
         \caption{}
     \end{subfigure}
     \vfill
    \begin{subfigure}[b]{0.2\textwidth}
         \centering
         \includegraphics[width=\textwidth]{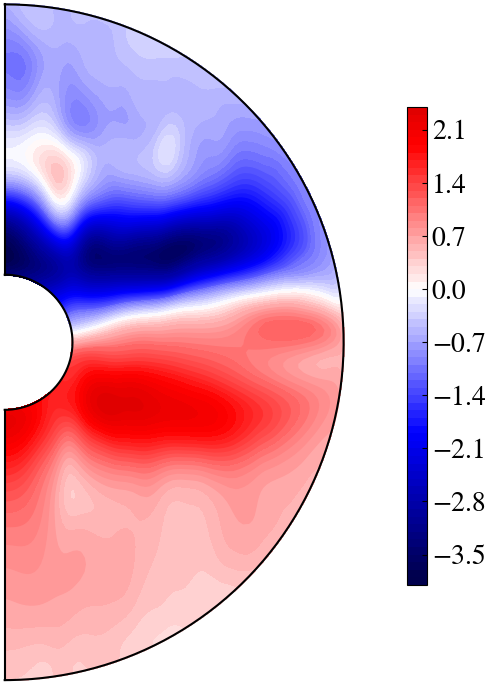}
         \caption{}
     \end{subfigure}
          \hfill
     \begin{subfigure}[b]{0.2\textwidth}
         \centering
         \includegraphics[width=\textwidth]{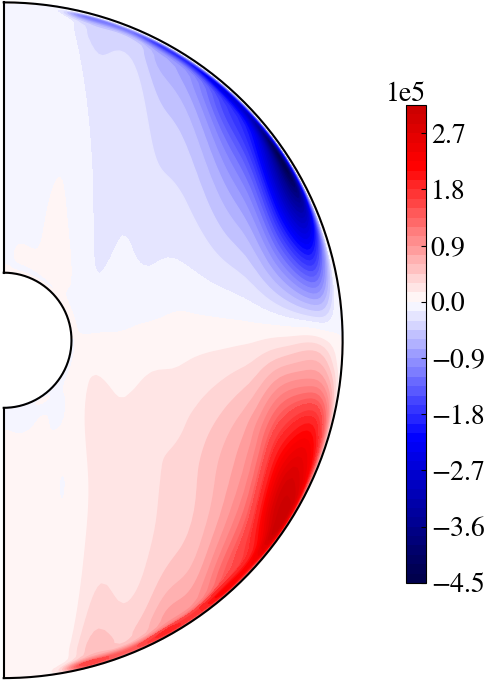}
         \caption{}
     \end{subfigure}
          \hfill
          \begin{subfigure}[b]{0.22\textwidth}
         \centering
         \includegraphics[width=1.04\textwidth]{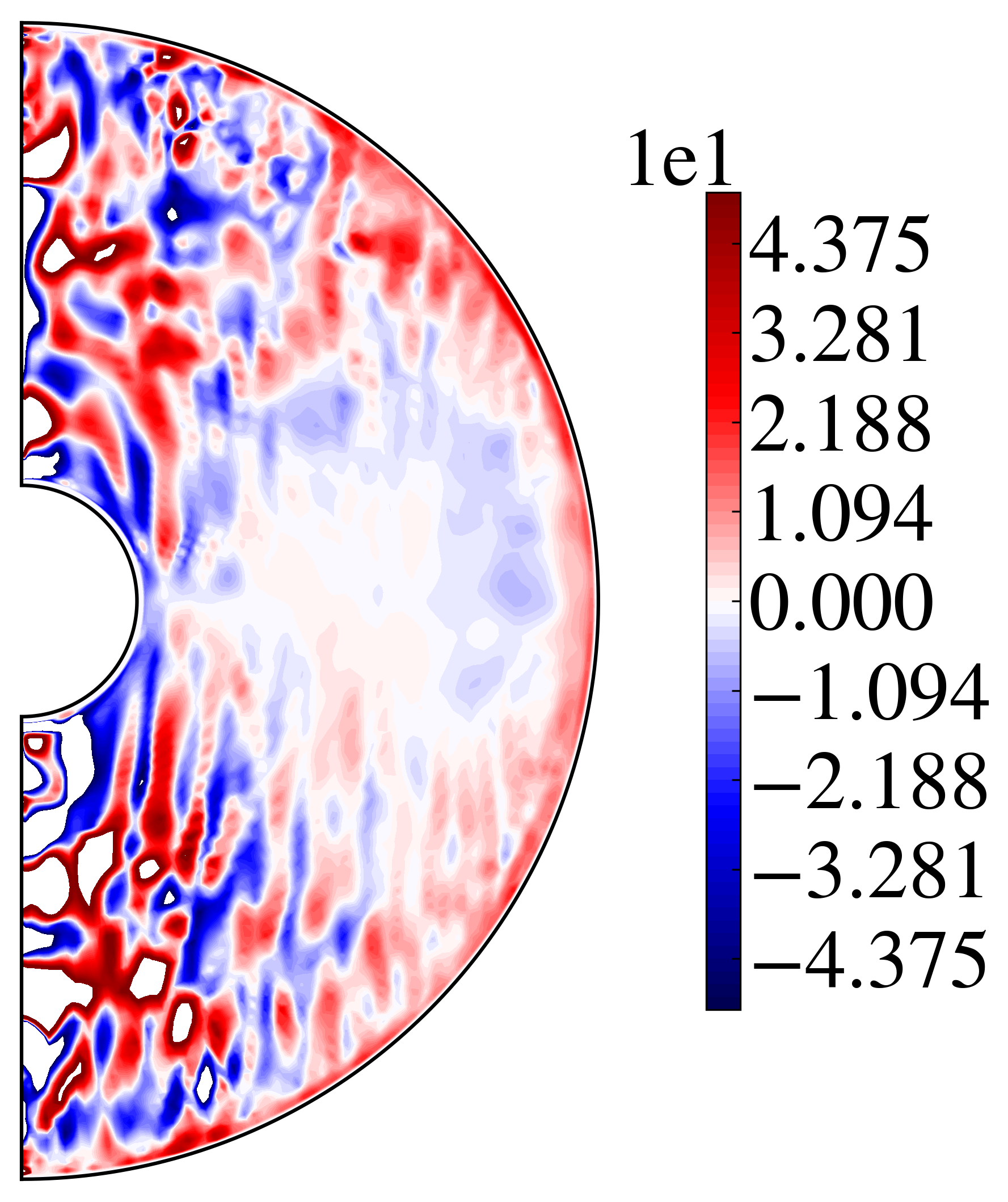}
         \caption{}
     \end{subfigure}
             \hfill
               \begin{subfigure}[b]{0.22\textwidth}
         \centering
         \includegraphics[width=0.94\textwidth]{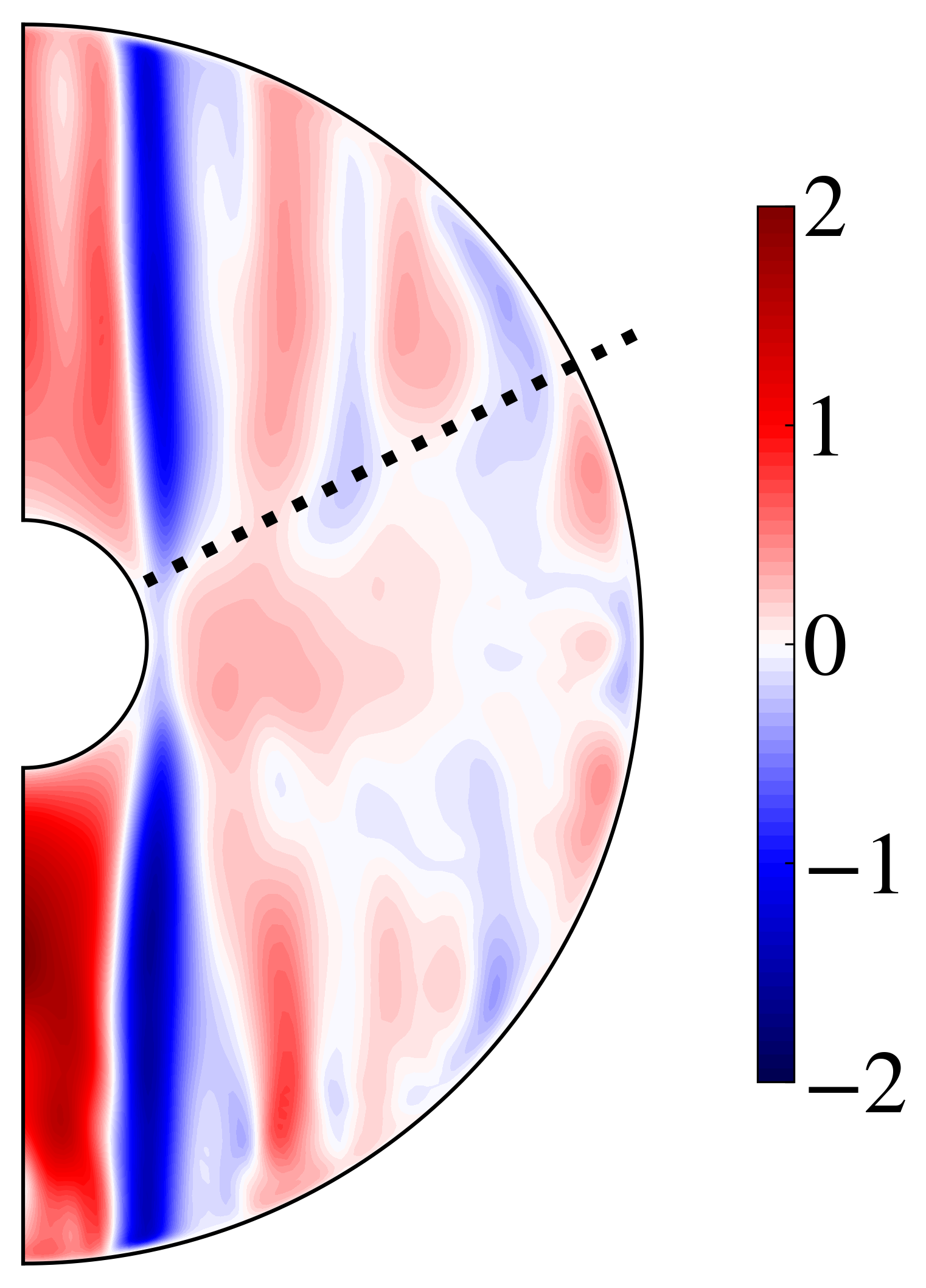}
         \caption{}
     \end{subfigure}
          \vfill

\begin{subfigure}[b]{0.2\textwidth}
         \centering
         \includegraphics[width=\textwidth]{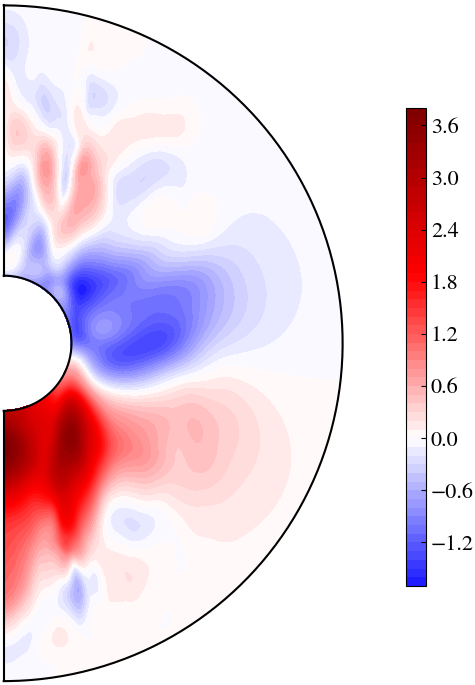}
         \caption{}
     \end{subfigure}
\hfill
     \begin{subfigure}[b]{0.2\textwidth}
         \centering
         \includegraphics[width=\textwidth]{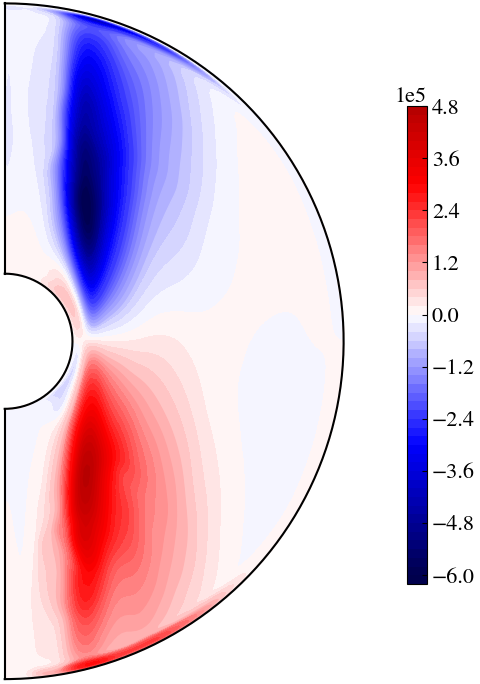}
         \caption{}
     \end{subfigure}
               \hfill
               \begin{subfigure}[b]{0.2\textwidth}
         \centering
         \includegraphics[width=\textwidth]{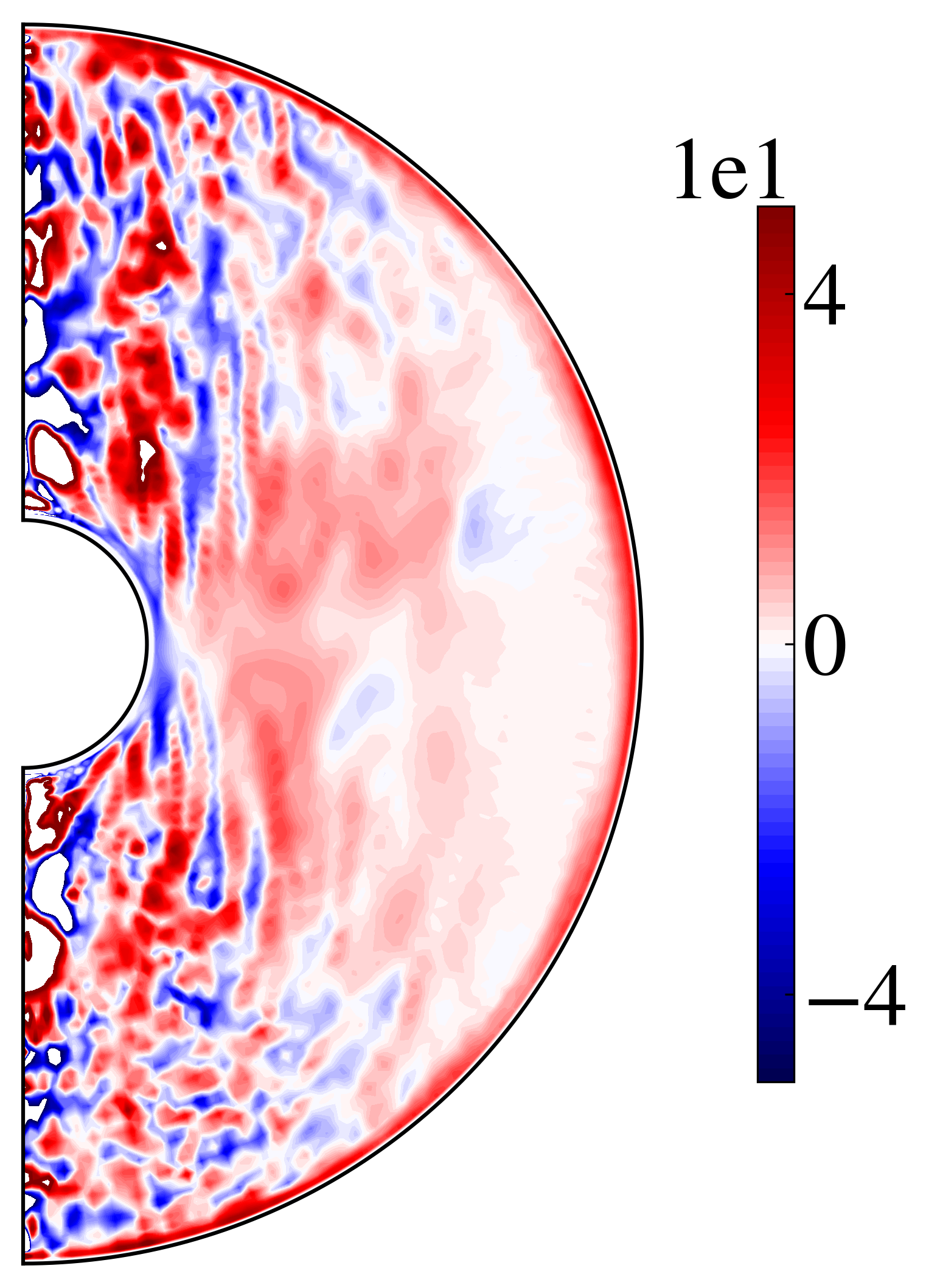}
         \caption{}
     \end{subfigure}
               \hfill
               \begin{subfigure}[b]{0.22\textwidth}
         \centering
         \includegraphics[width=0.94\textwidth]{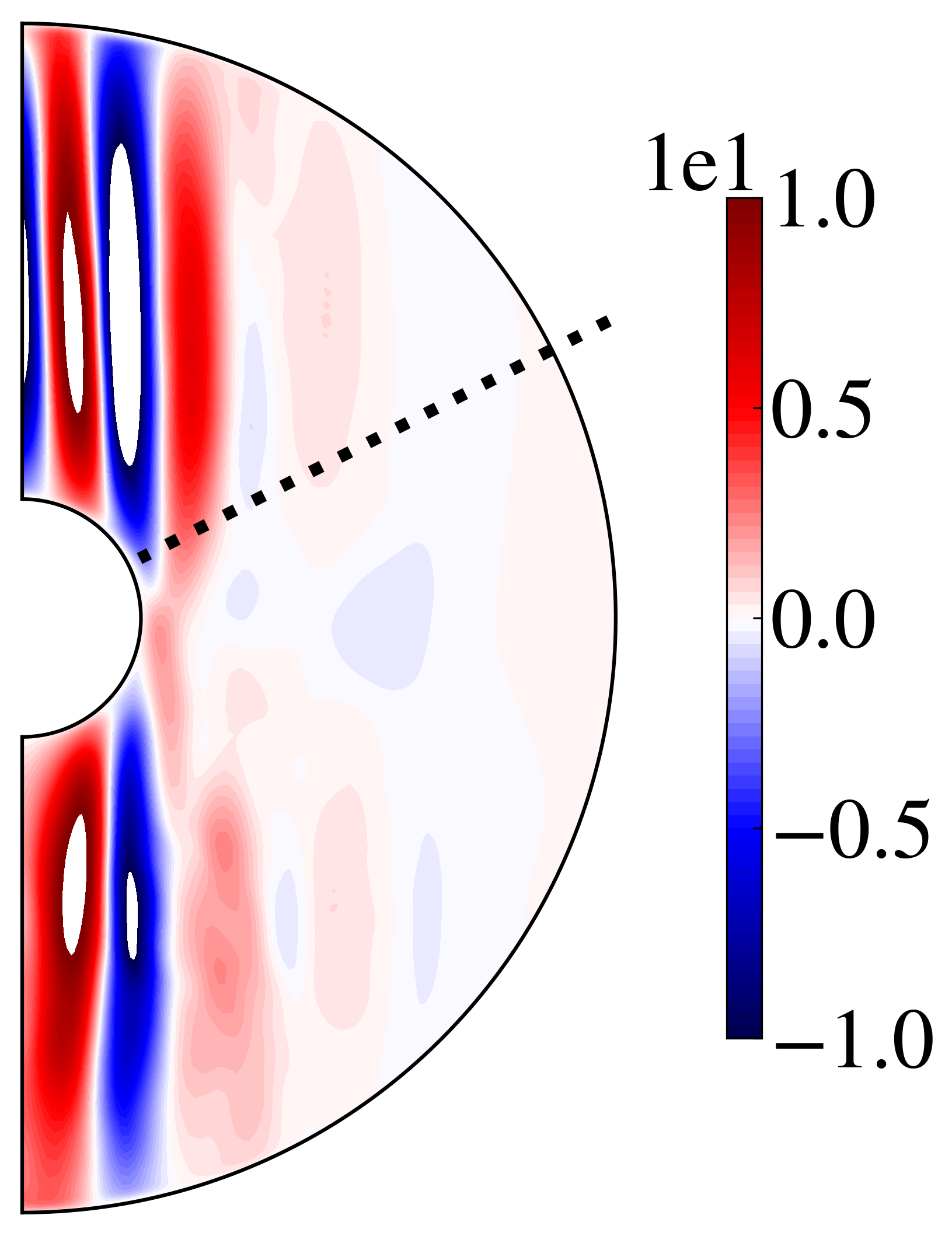}
         \caption{}
     \end{subfigure}
 \caption{From the left to the right: azimuthally averaged snapshots of radial magnetic field component $\langle B_r \rangle$, flow helicity $H = \langle \mathbf{U} \cdot \nabla \times \mathbf{U} \rangle$, the $r$-component of the $\gamma$-effect, \anna{axisymmetric by definition~\eqref{eq:mean_field_B}}, meridional circulation $\langle U_r\rangle$. \anna{All runs are without imposed differential rotation at the bottom of the convective zone} ($\Ross_{sh}=0$).  Top row (panels a-d): $g\propto 1/r^2$, $\Rayl=5\times10^6$, $\xi=0.35$; middle row (panels e-h): $g \propto r$, $\Rayl=8\times10^6$, $\xi=0.2$; bottom row (panels i-l): $g \propto 1/r^2$, $\Rayl=1\times10^6$, $\xi=0.2$. The dotted line corresponds to the considered latitude in the bottom panels of Figs.~\ref{fig:gamma_vs_Ur} and~\ref{fig:gamma_vs_Ur2}. \anna{The $\gamma$-effect and $\langle U_r \rangle$ are normalized using viscous units of the code (see Fig.~\ref{fig:cesam}); helicity is thus normalized by $\nu^2/d^3$. Magnetic field units are the same as in Fig.~\ref{fig:fdip_time}.}}
    \label{fig:slices}
\end{figure*}

\section{Mechanisms of dipole collapse}
\label{section:mechanism_collapse}
We interpret the mechanisms of the destabilization of dipoles in our simulations from the viewpoint of dynamo theory. The broken symmetry of rotating convective turbulence
is expected to drive axisymmetric fields due to non-zero correlation between flow velocity and vorticity ($\alpha$-effect). In addition, high anisotropy of convective structures leads to systematic advection redistributing magnetic field across the domain ($\gamma$-effect). Differential rotation,  developing due to turbulent momentum mixing, stretches poloidal field lines intro toroidal ($\omega$-effect). All these processes, opposed by turbulent diffusion $\eta_T$, support large-scale magnetic fields, according to the mean-field dynamo equations
\begin{equation}\label{eq:mean_field_B}
 \frac{\partial   \mathbf{\langle \bm{B} \rangle} }{\partial t}  = \nabla \times \left(  \left[\langle \bm{U} \rangle  + \bm{\gamma} \right] \times \langle \bm{B} \rangle +  \alpha \langle \mathbf{B} \rangle - \eta_T \nabla \times \langle \bm{B} \rangle \right).
\end{equation}
Vector $\bm{\gamma}$ can be interpreted as  additional, \anna{axisymmetric} mean-field advection velocity in Eq.~\eqref{eq:mean_field_B}, in addition to the \anna{axisymmetric} mean flow $\langle \bm{U}\rangle$. The final magnetic configuration will depend on the balance between these processes. Vector $\bm{\gamma}$ and tensor $\alpha$ can be computed directly from 3D data using SVD inversion method of~\cite{simard2016characterisation}, as it was done in this work.

\begin{figure}[h!]
    \centering
\begin{subfigure}[b]{0.47\textwidth}
         \centering
         \includegraphics[width=\textwidth]{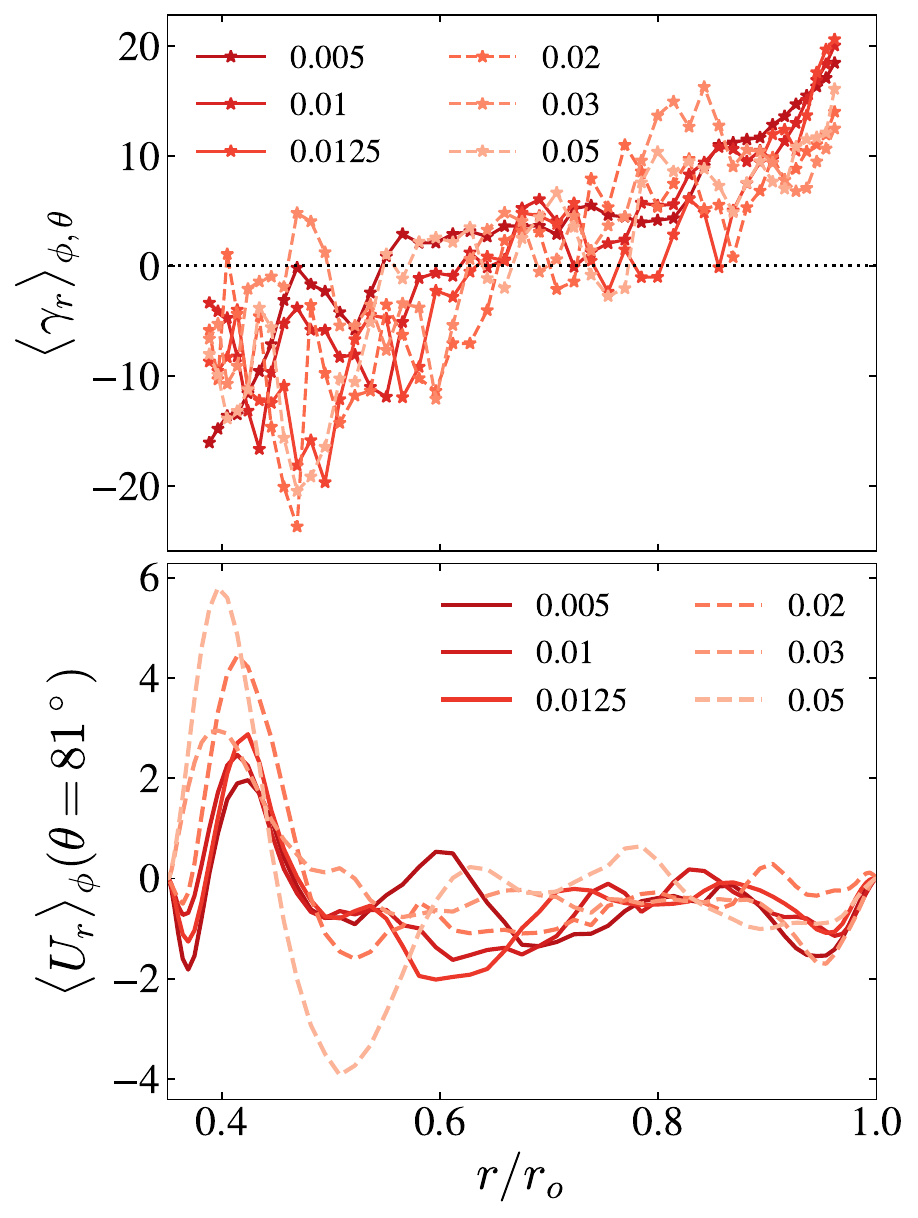}
     \end{subfigure}

    \caption{Top: radial component of the gamma-effect $\gamma_r$ as a function of $Ro_{sh}$ (legend), averaged over colatitude $\theta$. Bottom: radial component of meridional circulation, near the equator (colatitude of $\theta=80^\circ$, see Fig.~\ref{fig:slices}d). Solid lines, the dipole is stable,  dashed lines, the dipole collapsed. Runs with $g\propto 1/r^2$, $\Rayl=5\times10^6$, $\xi=0.35$ (\texttt{gr2\_2} in~Table~\ref{tab:sim_param}). \anna{Subscripts $\phi$ and $\theta$ denote azimuthal and latitudinal average, respectively. Same units as in Fig.~\ref{fig:cesam}(c-e).}}
    \label{fig:gamma_vs_Ur}
\end{figure}

\subsection{Induced $\omega$-effect}

Guided by general transitions of dipolar dynamos to oscillatory solutions, the need of high differential rotation and corresponding radial shear in Fig.~\ref{fig:Rosh_Roconv_local} to trigger dipole collapse hints 
towards a transition from $\alpha^2$ to $\alpha^2 \omega$ dynamos and excitation of oscillatory dipolar branches. Such transitions are typically accompanied by an increase in the toroidal component of magnetic field, which is stretched out by the differential rotation. To test this hypothesis, we plot the ratio between the rms values of the axisymmetric toroidal and poloidal components of magnetic field, integrated over the entire domain, in Fig.~\ref{fig:ro_sh_ro_conv}(c). For most of the models, this ratio remains relatively constant for stable dipolar solutions at small $\Ross_{\rm sh}$. As a general trend, it gradually increases as soon as periodic dipolar waves develop. This increase is stronger for thinner convective zones ($\xi=0.35$) and less pronounced for thicker ones ($\xi =0.1$), where wave onset is delayed. The fact that this increase is slow, and not abrupt, hints at gradual, supercritical transition between steady and oscillatory dipolar dynamo branches during dipole collapse. In contrast to that, the transition to quadrupoles at $g\propto r$ and high $\Rayl$ number is abrupt and thus potentially subcritical. We plan to investigate the bifurcation scenarios of these dynamo branches in more details in the future work.


\subsection{Disruption of $\gamma$-effect}
In the simulations with
$\xi=0.35$, $g\propto 1/r^2$, and with gravity profile in Eq.~(\ref{eq:g_r_cesam}) from \cesamxx, the dynamo action takes place predominantly outside the tangent cylinder and occupies nearly homogeneously the bulk of the flow (Fig.~\ref{fig:slices}b). Although the peak of magnetic energy is located inside the tangent cylinder, the large-scale dipole is occupying the entire domain (Fig.~\ref{fig:slices}a). As
$\Ross_{\rm sh}$ increases, we did not find a strong, systematic modification in helicity distribution or radial dependence of $\Ross_{\rm conv}$ number (Fig.~\ref{fig:Roconv_varom}), thus, imposed levels of rotation do not disturb significantly the length scale and magnitude of convective columns and generated by them $\alpha$-effect. Similarly for $\gamma$-effect, it remained nearly unmodified in this set of simulations, with the radial component $\gamma_r$ showing inwards pumping of magnetic energy in the bulk of the flow, and the outward direction of this component at the outer sphere. This distribution of $\gamma_r$ has been shown to have an important role in supporting dipolar dynamos, by re-distributing magnetic flux from the areas where toroidal field is generated, to the areas of poloidal field, providing additional exchange for  the two field components \citep{schrinner2012dipole}. When $\gamma_r$ is small,  dipolar dynamos ceased to exist and gave way to multipolar periodic fields. Decrease in efficiency of the $\gamma$-effect due to differential rotation could be one of the causes of the dipole collapse. However, our simulations do not exhibit strong variations in $\gamma_r$ as a function of $\Ross_{\rm sh}$, as shown in Fig.~\ref{fig:gamma_vs_Ur} (top panel). There,  the radial distribution of $\gamma_r$, integrated over latitude $\theta$, remains predominantly negative at lower radii and predominantly positive at higher radii, for both dipolar and non-dipolar runs.

On the other hand, due to mass and momentum conservation, imposed differential rotation leads to enhancement of poloidal (meridional) mean flows. Although in our simulations they have quite complex, multi-cellar topology corresponding to at least two large recirculation zones, all the runs feature an area of positive mean radial velocity  $\langle U_r \rangle_\phi$ (Fig.~\ref{fig:slices}d) developing near the equator and tangent cylinder. As $\Ross_{\rm sh}$ increases, the strength of this radial velocity becomes progressively larger (Fig~\ref{fig:gamma_vs_Ur}b). This velocity component, as seen from Eq.~\eqref{eq:mean_field_B}, is thus opposing negative magnetic pumping in the bulk of the domain and contributes to destabilization of the dipolar field component.

\subsection{Influence of the gravity profile}\label{sec:mechsm_infl_g}

The homogeneous profile $g\sim r$ results in mean magnetic field concentrated around the equator (Fig.~\ref{fig:slices}e). Intensity of convective columns is the largest near the outer boundary (Fig.~\ref{fig:cesam}d), which implies the concentration of flow helicity $H = \langle \mathbf{U} \cdot \nabla \times \mathbf{U} \rangle$ in the bulk at large radii (Fig.~\ref{fig:slices}f). The generation of magnetic field is thus located in this narrow zone, which has to be disturbed to obtain dipole collapse. Thus, the empirical local criteria for relation between the shear and convective Rossby numbers, like those in Fig.~\ref{fig:Rosh_Roconv_theta}, are satisfied in the outer layers ($r/r_{\rm o} \approx 0.9$), and not in the bulk of the flow. On the other hand, the relative topology of the $\gamma$-effect, although weakened in comparison to $g\sim 1/r^2$ profile, remains similar, with an area of negative radial mean flux of magnetic energy (Fig.~\ref{fig:slices}g), and strengthening of the meridional flow at the tangent cylinder in comparison to $\gamma_r$ (Fig.~\ref{fig:gamma_vs_Ur2}a). 
Our results suggest that this gravity profile exhibits delay in dipole collapse due to location of the bulk of helicity outside of the zone of strong shear. For higher $\Rayl$ number, the dipole is overstable, and drops on a entirely different, equatorially symmetric dynamo branch (Fig.~\ref{fig:fdip_time}d). We leave detailed study of stability of this gravity profile for the future work. Since it is relevant for stars just at the beginning of their formation and not on the PMS, we do not take it into account when deriving a more general shear criterion for dipole collapse on the PMS.

\begin{figure}
    \centering
    \includegraphics[width=\linewidth]{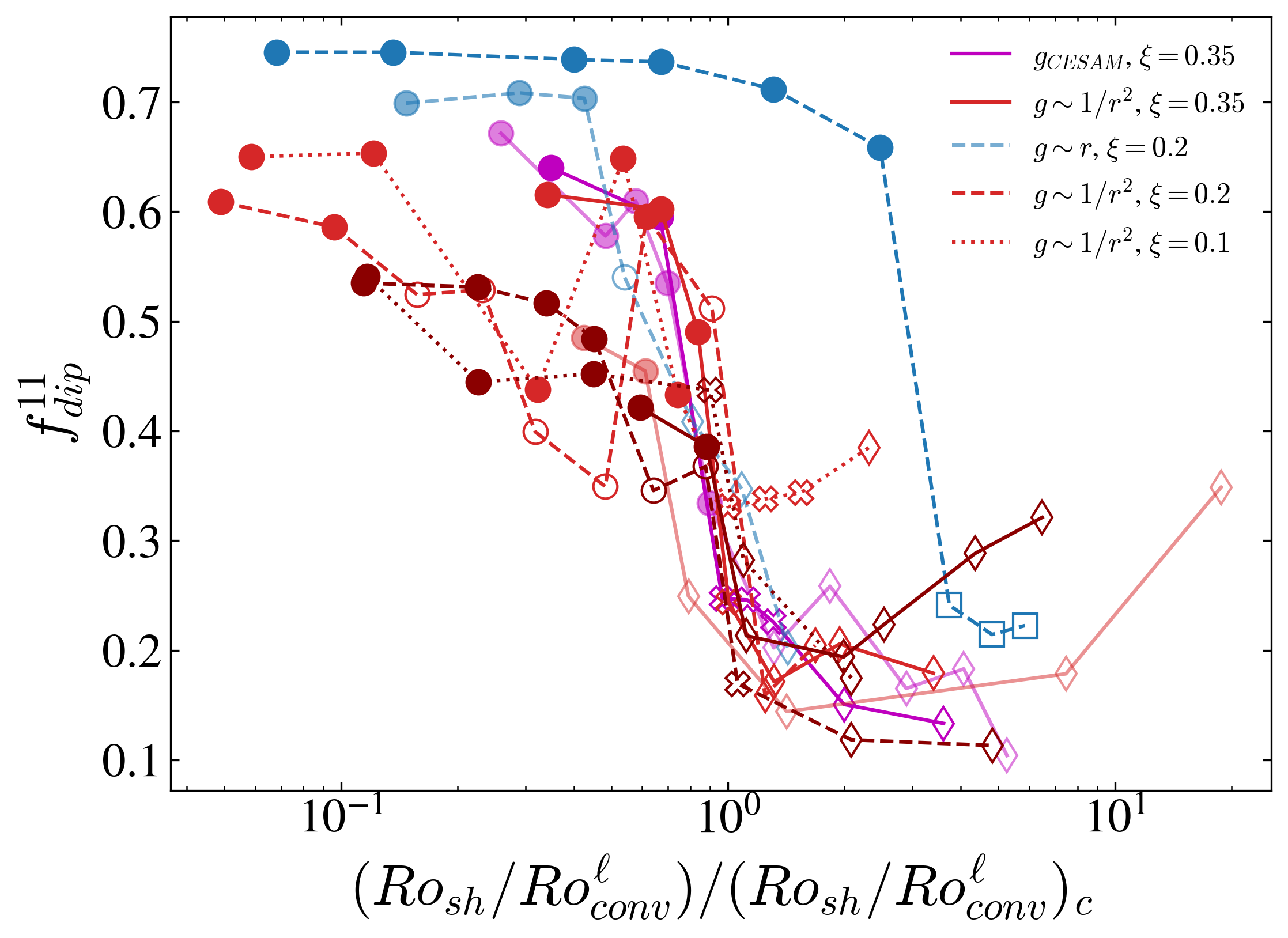}
    \caption{The data for time-averaged dipolarity from all runs of this paper, as a function of compensated shear Rossby number, i.e., $\Ross_{\rm sh}/\Ross_{\rm conv}^\ell$ normalized by its critical value given in Eq.~\ref{eq:gu_fit_form}. All transitions from dipolar to oscillatory branches collapse on each other. The branch that does not follow the trend is the special  case of $g\propto r$ and high $\Rayl$ values (run \texttt{gr\_2} in Table~\ref{tab:sim_param}, see Fig.~\ref{fig:fdip_time}d and discussion in Sect.~\ref{sec:mechsm_infl_g}). 
    Colors, symbols and lines as in Fig.~\ref{fig:ro_sh_ro_conv}.}
    \label{fig:fdip_compensated}
\end{figure}

\begin{figure*}[h!]
         \centering
         \includegraphics[width=\textwidth]{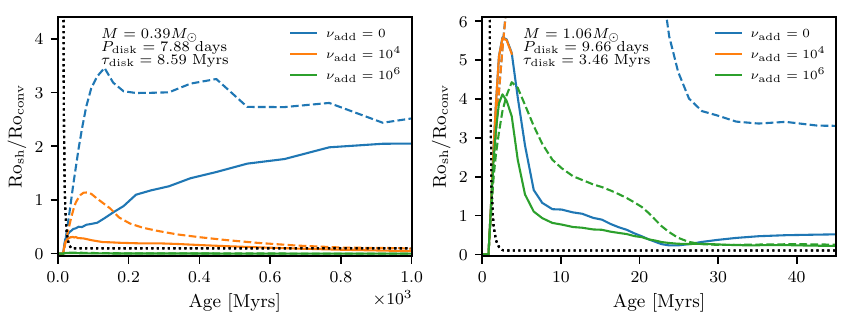}
    \caption{Evolution with stellar age of the ratios $\Ross_{{\rm sh}, H_p} / \Ross_{\rm conv, mid}$ (solid lines) or $\Ross_{\rm sh, glob} / \Ross_{\rm conv, mid}$ (dashed lines). The color corresponds to the additional viscosity considered for the angular momentum transport. The black dotted line corresponds to the threshold defined in Eq.~\eqref{eq:gu_fit_form} for $\Pram=4$, with a minimum value of 0.1. Each panel corresponds to a given evolutionary track, whose initial parameters are given in each plots.}
    \label{fig:Gu_evol}
\end{figure*}

\subsection{Impact of convective zone thickness}

In simulations with wider convective shells, $\xi=0.1$ or $0.2$, the bulk of helicity is located in deeper layers than for $\xi=0.35$ (compare Figs.~\ref{fig:slices}j and b), following the location of convective columns (Fig.~\ref{fig:cesam}e). The dynamo is thus ``deep-seated" at the bottom of convective zone, with magnetic energy decaying with radius (Fig.~\ref{fig:slices}i).  The magnetic fields of these dynamo solutions are ``weaker" and typically have lower Elsasser numbers (see Table~\ref{tab:sim_param}); stronger fields at higher levels of turbulence result multipolar in this parameter regimes, even without imposed radial differential rotation. In addition, the topology of the radial $\gamma$-effect considerably changes, developing weak but predominantly positive advection in the equatorial regions, repelling magnetic flux outwards. At high latitudes, the distribution of $\gamma_r$ appears noisy and small-scale (Fig.~\ref{fig:slices}k), and the radial direction of $\theta$-integrated magnetic pumping (Fig.~\ref{fig:gamma_vs_Ur2}b) is overall neither negative nor positive. However, the  latitudinal component $\gamma_t$ shows the same topology as in previous cases (compare panels a, b and c in Fig.~\ref{fig:gamma_t}), but shifted inwards, like the distribution of helicity. This indicates that magnetic flux is still potentially generated and  transported inwards at high latitudes inside the tangent cylinder, even if this is not obvious in Figs.~\ref{fig:slices} and ~\ref{fig:gamma_vs_Ur2} due to the noise. The area of magnetic pumping is thus partially bypassing the region of negative shear, developing due to rotation (like that in Fig.~\ref{fig:Rosh_Roconv_local}b-d). Thus, although meridional circulation is also enhanced at the tangent cylinder (Fig.~\ref{fig:gamma_vs_Ur2}b, bottom), it affects less the mean-field dynamo process and therefore dipole collapse is delayed, resulting in dipole reversals first. In addition, the location of negative shear region in the middle of dynamo-generating zone implies that the product of $\alpha \nabla \Omega$, defining the direction of the waves~\citep{yoshimura1975solar,tobias2021turbulent}, changes sign across the dynamo-generating area. This may explain why the dynamo falls first on aperiodic solutions without well-defined frequency at  $\xi = 0.1$ (Fig.~\ref{fig:ro_sh_ro_conv}a,b), before developing oscillating dynamos.

\section{Discussion on stellar applications }
\label{section:discussion}

\subsection{Fitting the dipole collapse threshold in DNS}
\label{sect:fit coll}

For practical stellar applications, it can be useful to obtain a simple ad hoc expression for the critical value of $\Ross_{\rm sh}/\Ross_{\rm conv}^\ell$ above which dipoles collapse as a function of the aspect ratio $\xi$. Indeed, $\xi$ is representative of the internal structure of PMS stars and globally increases with age (see Sect.~\ref{section:early_stell_evol}). To derive such an expression, we estimate the values of $\Ross_{\rm sh}/\Ross_{\rm conv}^\ell$ before and after collapse at a given $\xi$ value, and compute the mean value. In Fig.~\ref{fig:ro_sh_ro_conv}d we plot these mean values as a function of $\xi = r_{\rm i} /r_{\rm o}$, for $\Pram=4$ and $6$. The solid lines correspond to exponential fits to these critical mean thresholds in the form of $C(\Pram) \exp( -\beta \xi)$,
\begin{equation}
   \left( \frac{\Ross_{\rm sh}}{\Ross_{\rm conv}^\ell} \right)_{\rm c} \approx \left\{
    \begin{array}{ll}
        25.61 \exp(-15.47 \xi) \quad \text{for}~\Pram=4 \\[2pt]
          11.44 \exp(-15.36 \xi) \quad \text{for}~\Pram=6,
\label{eq:gu_fit_form}
    \end{array}
\right.  
\end{equation}
The dependence on $\xi$ in the exponent function is the same for both considered $\Pram$ values.
Varying $\Pram$ modifies only the pre-factor. 
While for thick shells, $\xi=0.1$, the dipole is two times more  stable at $\Pram=4$, this difference becomes smaller as $\xi$ increases, and  nearly negligible around $\xi =0.35$.
The last criterion thus tends to become independent on $\Pram$ around $\xi \sim 0.35$, that is, the dipole stability is expected to be less and less dependent on the relative strength of stellar magnetic field (increasing with $\Pram$). The collapse threshold $(\Ross_{\rm sh}/\Ross_{\rm conv}^\ell )_{\rm c}\sim 0.1$, valid for $\xi =0.35$, can thus be seen as a limiting ratio - all stars with convective zones thinner than this one can be supposed to have their dipoles collapsed if $\Ross_{\rm sh}/\Ross_{\rm conv} ^\ell\gtrsim 0.1$. This will be assumed in Sect.~\ref{sect:evoldip}. We plot in Fig.~\ref{fig:fdip_compensated} the dipolarity as a function $\Ross_{\rm sh}/\Ross_{\rm conv}^\ell$ normalized by its critical value taken from Eq.~\eqref{eq:gu_fit_form}; this perfectly illustrates the shear-induced dipole collapse identified in this work.

\subsection{
Estimate of the Rossby numbers in 1D stellar evolution models}
\label{sect:adapt3d1d}

The expression of the shear Rossby number $\Ross_{\rm sh}$ in Eq. \eqref{eq:rosh} is not straightforward to translate into 1D stellar models. First, the radial shear $\Delta\Omega$ entering $\Ross_{\rm sh}$ has no equivalent in a \cesamxx model as the convective zone is assumed to rotate as a solid body. 
This is nevertheless certainly not true in stars, where a degree of shear should exist close to the boundary with the radiative core (see Sect.~\ref{section:early_stell_evol}). Therefore, we estimated, on the one hand, an inner angular velocity $\Omega_{\rm i}$ as the angular velocity in the radiative zone (RZ) averaged with its moment of inertia:
\begin{equation}
    \Omega_{\rm i, glob} = \frac{\int_{\rm RZ} r^2 \Omega{\rm d} m}{\int_{\rm RZ} r^2{\rm d} m}.
\end{equation}
On the other hand, one may argue that the effect of the angular velocity of the very deep regions of the RZ on the stability of the dipole is negligible compared to that of the angular velocity close to the convective boundary. Then, it seems reasonable to consider the value of $\Omega$ at a pressure scale height below the bottom of the convective zone $r_{\rm CZ} - H_p$, which defines $\Omega_{{\rm i}, H_p}$.

Second, for the convective Rossby number  $\Ross_{\rm conv}^\ell$, the radial mean of Eq.~\eqref{eq:roc}, can be estimated  in evolution models using the MLT convective velocity, $v_{\rm conv}$. 
Provided the global turnover timescale is $\tau_{\rm conv} =  \int_{\rm CZ}{\rm d}r/v_{\rm conv}$, it is straightforward to define a first global quantity $\Ross_{\rm conv, glob} = 1 / (\Omega_{\rm CZ}\tau_{\rm conv})$.  MLT, however, fails to describe the very near surface layers which could bias this estimate. We thus prefer in the following to consider $\Ross_{\rm conv, mid} = v_{\rm conv}(r_{\rm mid}) / [\Omega_{\rm CZ} l_{\rm MLT} (r_{\rm mid})]$, computed at the middle of the convective zone $r= r_{\rm mid}$, with the associated mixing length $l_{\rm MLT}$.

With these definitions, one can compute two different estimates of $\Ross_{\rm sh} / \Ross_{\rm conv}^\ell$, which are more or less favorable to the stability of the dipole. Figure \ref{fig:Gu_evol} shows variation of some of these ratios for few evolutionary tracks. As the current description of angular momentum transport notoriously lack efficiency, we have considered for each track an additional turbulent viscosity $\nu_{\rm add} = 0$, $10^4$ and $10^6~\textrm{cm}^2~\textrm{s}^{-1}$ in the radiative layers. This maximum value is compatible with the potential, still debated, transport by MHD instabilities \citep[e.g.][for a solar model]{Eggenberger2005}. We see that $\Ross_{\rm sh,glob} / \Ross_{\rm conv,mid}$ is generally much larger than $\Ross_{{\rm sh},H_p} / \Ross_{\rm conv,mid}$, as expected since the fast core rotation contributes to the value of $\Ross_{\rm sh,glob}$ positively. This difference decreases as $\nu_{\rm add}$ increases, since the rotation profile gets smoother and smoother. In both cases, the dipole collapse criterion provided by Eq.~\eqref{eq:gu_fit_form} for $\Pram=4$ is met at the very beginning of the PMS evolution (i.e., around 1 Myr) if $\nu_{\rm add} $ is below a threshold value: the lower the mass, the lower the threshold value for $\nu_{\rm add}$.
We suggest that the ratio $\Ross_{{\rm sh}, H_p} / \Ross_{\rm conv, mid}$, based on the rotation in upper layers of radiative zone adjacent to CZ, is the more consistent with the results of the 3D simulations, where $\Delta \Omega$ is imposed at the bottom of the CZ.  Moreover, if its value is higher than the critical value for dipole collapse, this is also the case for $\Ross_{\rm sh,glob} / \Ross_{\rm conv}$, ensuring a conservative approach.  

\begin{figure}
    \centering
    \includegraphics[width=\linewidth]{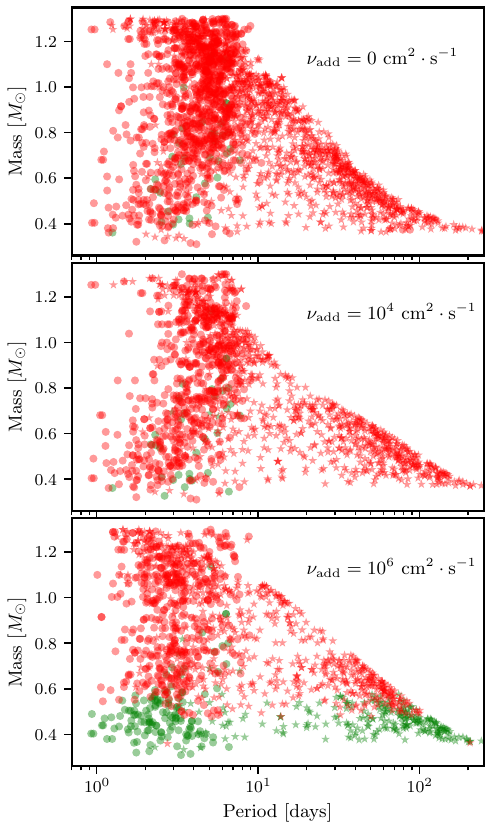}
    \caption{Mass-Period (surface) distribution of some models randomly chosen in the grids described in Sect. \ref{sect:evoldip}. The criterion considered for the stability of the dipole is the ratio $\Ross_{{\rm sh}, H_p} / \Ross_{\rm conv, mid}$ with the threshold expressed in Eq. \eqref{eq:gu_fit_form}. Circles and stars are respectively PMS and MS models. Red (resp. green) markers are models that lost (resp. kept) their dipole between their formation and their present age. Each panel corresponds to a grid with a given additional turbulent viscosity $\nu_{\rm add}$.}
    \label{fig:donati_plots_past}
\end{figure}

\subsection{Dipole stability through evolution}
\label{sect:evoldip}

Using the parameter $\Ross_{{\rm sh}, H_p} / \Ross_{\rm conv, mid}$ and  the criterion in Eq. \eqref{eq:gu_fit_form}, we can explore the stability of an initial dipole during the PMS. To that end, we construct three grids (for $\nu_{\rm add} = 0$, $10^4$ and $10^6~\textrm{cm}^2~\textrm{s}^{-1}$, see Sect.~\ref{sect:adapt3d1d}) of 700 stellar evolutionary tracks using the code \cesamxx (see Sect. \ref{subsection:cesam}), with different initial conditions. \lm{hence, one track corresponds to a series of structure models with same initial mass, $P_{\rm disk}$ and $\tau_{\rm disk}$, but different ages.} Tracks in each grid have a mass between $0.3$ and $1.3 M_\odot$, $P_{\rm disk}$ between $6$ and $10$ days and $\tau_{\rm disk}$ between $1$ and $10$ Myrs, chosen following a Sobol' pseudo-random algorithm \citep{Sobol1967,Bratley1988}. This choice avoids spurious patterns appearing in the final distributions. Because we are interested in early phases, each track is started in the CTTS phase, until the beginning of the MS phase (when $10\%$ of the initial hydrogen content is consumed). \lm{An illustration of the evolution of the surface rotation period of these models can be found in App. \ref{app:prot_evol}, Fig. \ref{fig:prot_evol}.}

We then compare $\Ross_{{\rm sh}, H_p} / \Ross_{\rm conv, mid}$ to the critical threshold for dipole collapse in Eq. \eqref{eq:gu_fit_form} (for $\Pram=4$ and a minimum value equal to $0.1$) over the stellar models grids. For each of the grids, we choose randomly 5 models per track. This is done to improve the readability of the distribution, and does not affect our conclusion thanks to the still large number of models considered. Fully convective models, i.e., $M\lesssim 0.35M_\odot$, are not considered. We also reject models with an age below $\tau_{\rm disk}$ because that may introduce spurious patterns in our distributions (i.e., two models of the same track \lm{still coupled to the disk will be at the same place in the mass-period diagram}). Figure \ref{fig:donati_plots_past} represents in red the models that have lost their dipole in their past life and in green, the one that maintained it. These distributions are mainly bi-modal with a group for PMS (small periods; circles) and a group for the MS stars (long period; stars). 
Whatever the value of the additional viscosity is, there are few models that keeps their initially stable dipolar field. Their number increases when $\nu_{\rm add}$ increases, incorporating older and older models. For $\nu_{\rm add} \lesssim 10^4~\textrm{cm}^2~\textrm{s}^{-1}$, few stable dipoles are found almost only in PMS models, while when $\nu_{\rm add} \gtrsim 10^6 ~\textrm{cm}^2~\textrm{s}^{-1}$ many MS models have managed to keep it. We also notice a strong correlation with the mass of the model. The more massive models generally have a stronger increase of the angular velocity during initial contraction than less massive ones, which results in a higher $\Ross_{{\rm sh}, H_p}$ (see Fig.~\ref{fig:Gu_evol}).  

In case $\nu_{\rm add}\gtrsim 10^6~\textrm{cm}^2~\textrm{s}^{-1}$, the predicted distribution of magnetic topology with mass on the MS qualitatively reproduces the mean observed trends, with dipolar (mostly poloidal) state for masses $0.35 M_\odot \lesssim M \lesssim 0.5 M_\odot$ and a transition towards more complex (toroidal) configurations for more massive low-mass stars \cite[e.g.,][]{Donati2009,Moutou2017,Donati2023}. This suggests that the distribution of magnetic fields with mass on the MS for masses $M\gtrsim 0.35M_\odot$ is likely to result from a (still unknown) internal angular momentum redistribution process in the radiative core whose efficiency is such that the induced radial differential rotation can destabilize dipoles only for masses $M\gtrsim 0.5M_\odot$. Although less obvious, the distribution with mass on the PMS somehow shows similarities too \citep{Nicholson2021}. This observed behaviour is of course dependent of the way we model the accretion and PMS phases and on the assumptions made for the evolution of magnetic fields (here, the dipole is assumed to be lost forever after its collapse). Future studies are needed to tackle these questions, out of scope of this work.

\section{Conclusions}
\label{section:concl}

Large-scale differential rotation, developing in thick convective layers of PMS stars, can perturb the stability of their strong initial dipolar fields. In this work, we developed 3D numerical dynamo models of this process, with imposed shear rates not perturbing convection significantly and well below the threshold for shear instability, which could also affect the field. We show that the key parameter for the stability of dipolar dynamos of PMS stars is the relative importance of a shear-based Rossby number compared to the convective Rossby number (see Figs.~\ref{fig:ro_sh_ro_conv} and \ref{fig:fdip_compensated}). Surprisingly, even relatively weak values of this ratio can lead to dipole collapse.

We considered different, more or less realistic, gravity profiles in our 3D DNS affecting the location of turbulent motions (see Figs.~\ref{fig:grav_omega} and \ref{fig:cesam}), which could correspond to PMS stars at different ages. In general, the convective Rossby number is radially dependent (Fig.~\ref{fig:grav_omega}b) as observed in strongly stratified DNS \citep{Raynaud2015}. While it is instructive to explore local criteria for dipole stability (Fig.~\ref{fig:Rosh_Roconv_theta}), such criteria can not be used in 1D stellar evolution codes like \cesamxx, where the access to 2D distribution of shear and convection is not available. We thus developed a criterion for dipole collapse based on the radially averaged value of convective Rossby number, and the total shear applied to the convective zone (Eq.~\ref{eq:gu_fit_form}). Eventually, this ratio depends on the thickness of convective zone, converging to universal behavior as the star evolves and the zone becomes thinner (Fig.~\ref{fig:ro_sh_ro_conv}b). Our simulations with stronger magnetic effects (higher $Pm$) also require a slightly modified criteria \citep[see also][]{menu2020magnetic,Zaire2022}. However, a more extensive parameter study with higher $Pm$ is needed for exploring magnetostrophic regimes in which the Lorentz force enters the dominant force balance \citep{dormy2016strong} and for determining a relevant scaling law for the stability criteria as a function of magnetic field strength (or $Pm$). 

Close to the transition between dipolar and oscillatory dynamos, the differential rotation does not affect significantly the distribution of convective Rossby number (Fig.~\ref{fig:Roconv_varom}), and thus helicity distribution in the flow for varying $Ro_{\rm sh}$ (not shown here). Helicity is strongly linked with $\alpha$-effect in the dynamo theory, i.e. conversion from toroidal magnetic fields into large-scale poloidal ones, and so this result suggests that $\alpha$-effect is not affected by differential rotation. By contrast, increasing imposed shear promotes toroidal magnetic fields generated by differential stretching of poloidal field lines ($\omega$-effect, see Fig.~\ref{fig:ro_sh_ro_conv}c) and affects the $\gamma$-effect, transport mechanism of particular importance for dipolar dynamos \citep[see Figs.~\ref{fig:gamma_vs_Ur} and \ref{fig:gamma_t}, and][] {schrinner2012dipole}. The differential rotation governs the dynamo behaviour beyond the critical values for collapse, with consistent migration of radial fields towards equator in agreement with Parker theory, i.e. the propagation direction depends on the sign of the radial differential rotation \citep{parker1955hydromagnetic,schrinner11}. The appearance of more complex field morphologies and temporal evolution, like aperiodic reversals and quadrupoles (Fig.~\ref{fig:fdip_time}c,d), suggest potential bi-stability of different dynamo branches.

Finally, we adapted this criterion for 1D stellar evolution code \cesamxx, estimating the ratio of shear-based Rossby number and the convective one during the evolution of stars, as shown in Fig.~\ref{fig:Gu_evol}. We estimated whether the dipole collapsed or not during the PMS phase for random distributions of stellar models, constructed to mimic observations.  Our results show that the redistribution of angular momentum taking place during the PMS phase can destabilize dipolar fields inherited from the proto-stellar phase and give rise to the diversity of magnetic topologies observed for main-sequence stars (see Fig.~\ref{fig:donati_plots_past}), \anna{when strong turbulent viscosity in the radiative zone is considered}. \anna{This provides an additional hint for more efficient transport of angular momentum in radiative zones of stars, than provided by shear and meridional currents alone.} \anna{The Sun with its relatively large mass, moderate rotation and periodically flipping, oscillatory magnetic field qualitatively fits well  in the scenario of steady dipolar dynamo  collapse and transition to oscillatory dynamos, as observed in this work.}  Additional numerical studies are needed to determine if dipolar dynamos can be generated from oscillatory dynamos when shear flows decrease during the beginning of the MS phase. To extrapolate these results to more massive stars, more sophisticated 3D DNS with adjacent stable and thermally unstable regions must be dynamically modeled.

\begin{acknowledgements}
The authors acknowledge financial support from the French program ‘PROMETHEE’ (Protostellar Magnetism: Heritage vs Evolution) managed by Agence Nationale de la Recherche (ANR), and would like to thank E. Alecian and the PROMETHEE team for inspiring discussions that helped to shape this work. This work was supported by the “Action Thématique de Physique Stellaire” (ATPS) of CNRS/INSU PN Astro co-funded by CEA and CNES, and the HPC facility MesoPSL for the computational resources. L. M. acknowledge support from the Agence Nationale de la Recherche (ANR) grant ANR-21-CE31-0018
\end{acknowledgements}

\bibliographystyle{aa}
\bibliography{biblio}

\begin{appendix}
\onecolumn

\section{Physical ingredients of \cesamxx models}
\label{appendix:cesam_phy}
\FloatBarrier
{\renewcommand\arraystretch{1.8}
\begin{table*}[h!]
    \caption{Main physical ingredients used in the \cesamxx models computed for this work.}
    \begin{tabular}{L{0.3\textwidth} L{0.6\textwidth}}
         \hline\hline
         Initial chemical composition & \begin{minipage}[t]{0.6\textwidth}
                                        Reference solar chemical composition of \citet{Asplund2009}.
                                        \end{minipage}\\
         Convection                   & \begin{minipage}[t]{0.6\textwidth}
                                        Mixing-Length Theory formalism \citep{Bohm-Vitense1958,Henyey1965}. 
                                                 \end{minipage}\\
         Atmosphere                   & \begin{minipage}[t]{0.6\textwidth}
                                        Reconstructed from a Hopf $T(\tau)$ relation \citep{Hubeny2014}. 
                                                 \end{minipage}\\
         Opacity                      & \begin{minipage}[t]{0.6\textwidth}
                                        Opacity tables from the OPAL team \citep{Rogers1992,Iglesias1996,Rogers2002}, adapted to our choice of chemical composition and supplemented at low temperatures by the Wichita opacity tables \citep{Ferguson2005}. 
                                                 \end{minipage}\\
         Equation of state            & \begin{minipage}[t]{0.6\textwidth}
                                        Tabulated by the same collaboration \citep[OPAL5,][]{Rogers2002}. 
                                                 \end{minipage}\\
         Nuclear reaction rate        & \begin{minipage}[t]{0.6\textwidth}
                                        Compilations from NACRE \citep[except LUNA for the ${}^{14}{\rm N(p},\gamma){}^{15}{\rm O}$ reaction; \citealt{Broggini2018}]{Aikawa2006}.
                                                 \end{minipage}\\
         Transport of chemical elements & \begin{minipage}[t]{0.6\textwidth}
                                        Neglected. 
                                                 \end{minipage}\\
         Transport of angular momentum & \begin{minipage}[t]{0.6\textwidth}
                                        Prescription of \citep{Talon1997b} with shear-induced turbulence modeled following \citet{Mathis2018}. Convective zone are supposed to rotate as solid body.
                                                 \end{minipage}\\
         Angular momentum loss         & \begin{minipage}[t]{0.6\textwidth}
                                        Scaling relation of \citet{Matt2015}.
                                                 \end{minipage}\\
         \hline
    \end{tabular}
    \label{tab:cesam_param}
\end{table*}}

\FloatBarrier

\section{\anna{Parameters and initial conditions for 3D simulations}}\label{sec:appA}

\anna{DNS code MagIC uses a set of dimensionless parameters that allow to control physical properties of our simulations. The ratio of viscous to Coriolis forces is Ekman number $E=\nu/\Omega_s d^2 = 10^{-4}$, where $\nu$ is hydrodynamic viscosity, $d=r_o-r_i$ the gap between  the two spheres, and $\Omega_s$ is the angular velocity of the surface (i.e. of the rotating frame). Furthermore, the density contrast $N_\rho = \ln{\left(\rho_i/\rho_o\right)}=2.5$ allows to take into account the drop of density, $\rho$, across the convective zone.  Prandtl and magnetic Prandtl numbers, $Pr =\nu/\kappa =1$ and $Pm = \nu/\eta$, where $\eta$ is magnetic diffusivity and $\kappa$ thermal diffusivity, define diffusive properties of the fluid. Rayleigh number $Ra= \left(GM d\Delta S\right)/(\nu \kappa c_p)$, where  $c_p$ heat capacity, $G$ gravitational constant, and $M$ the central mass, controls the strength of convection; see Table~\ref{tab:sim_param} for $Pm$ and  $Ra$. Table~\ref{tab:sim_param} also shows diagnostic parameters that allow comparison of our simulations with existing literature. Among them are the Elsasser, Reynolds and Nusselt numbers,
\begin{equation}
 \Lambda = \frac{B_{rms}}{\mu_0 \eta \rho_o \Omega_s}, \quad Re = U_{rms} d/\nu, \quad   Nu = \frac{\int_S - \kappa \rho T \frac{\partial S}{\partial r} r^2 \sin{\theta} d\theta d\phi}{4 \pi n c_1 \zeta_i^n \left( \exp N_\rho -1 \right)^{-1} },
\end{equation}
where $U_{rms}$ and $B_{rms}$  are the root-mean-square velocity and magnetic field, $T$ is temperature, $S$ entropy, $c_1$ and $\zeta_i$ are functions of the polytropic reference state, and $\mu_0$ is the permeability of the free space. We refer the reader to the previous works (i.e. \cite{Guseva2025,jones2011anelastic,dormy2016strong}) for a detailed discussion of these parameters.}

\FloatBarrier
{\renewcommand\arraystretch{1.5}
\begin{table*}[h!]
    \centering
    \caption{Parameters of dipolar initial conditions used in this paper. The values of all parameters, except the last column, are given for the base case without differential rotation, $Ro_{\rm sh}=0$, and do not change considerably when $\Ross_{\rm sh} \ne 0$ (e.g., see~Fig.~\ref{fig:Roconv_varom}). The last column corresponds to the shear Rossby number at the dipole collapse, $\Ross^{\rm collapse}_{\rm sh}$, allowing to reconstruct entire simulation space. 
    }
    \begin{tabular}{c c c c c c c c c c }
    \hline\hline
        Label & $\Pram$  & $\xi$ & $\Rayl$   & $\Ross_{\rm conv}$  & $\Lambda$ & $f^{11}_{\rm dip}$ &  $\Reyn$ & $\Nuss$  &   $\Ross^{\rm collapse}_{\rm sh}$   \\ \hline
         \multicolumn{10}{c}{ $g \sim 1/r^2$}\\ 
         \verb|gr2_1|  & $4$& $0.35$ & $2 \cdot 10^6$ &  0.044& 9.35& 0.54& 67.9& 2.53& 0.00425 \\ 
         \verb|gr2_2|  & $4$& $0.35$ & $5 \cdot 10^6$ &  0.127& 29.58& 0.67& 206.2& 7.26&  0. 0135\\ 
         \verb|gr2_3|  & $4$& $0.2$ & $1 \cdot 10^6$  &  0.025& 1.68& 0.66& 52.6& 1.78&  0.035\\ 
         \verb|gr2_4|  & $4$& $0.1$ &  $5 \cdot 10^5$ &  0.015& 1.49& 0.67& 47.7& 2.07& 0.085 \\ 
         \verb|gr2_5|  & $6$& $0.35$ & $1.5 \cdot 10^6$ &  0.026& 9.75& 0.56& 42.7& 1.82&  0.00175\\ 
         \verb|gr2_6|  & $6$& $0.2$ &  $1.5 \cdot 10^6$ & 0.040& 19.93& 0.56& 94.8& 3.75&  0.0175\\ 
         \verb|gr2_7|  & $6$& $0.1$ & $5.5 \cdot 10^6$ &  0.016& 1.49& 0.67& 47.6& 2.07&  0.0355\\ \hline
        \multicolumn{10}{c}{ $g$, \cesamxx}\\ 
         \verb|gc_1|  & $4$ & $0.35$ & $1 \cdot 10^7$&  0.084& 20.06& 0.68& 118.5& 4.14&  0.00875\\ 
         \verb|gc_2|  & $4$ & $0.35$  & $1.5 \cdot 10^7$&  0.121& 31.42& 0.71& 171.1& 5.78& 0.0125 \\ \hline
         \multicolumn{10}{c}{ $g \sim r$}\\ 
         \verb|gr_1|  & $4$ & $0.2$ & $8 \cdot 10^6$ &  0.029& 1.44& 0.73& 41.8& 1.42& 0.025 \\ 
         \verb|gr_2|  & $4$ & $0.2$ & $1.6 \cdot 10^7$ &  0.062& 15.09& 0.74& 104.5& 2.98& 0.25 \\ 
    \end{tabular}
    \label{tab:sim_param}
\end{table*}}



\FloatBarrier


\section{Supplementary figures highlighting the mechanisms of dynamo collapse}

\FloatBarrier

\begin{figure*}[!h]
\centering
    \begin{subfigure}[b]{0.24\textwidth}
         \centering
         \includegraphics[width=\textwidth]{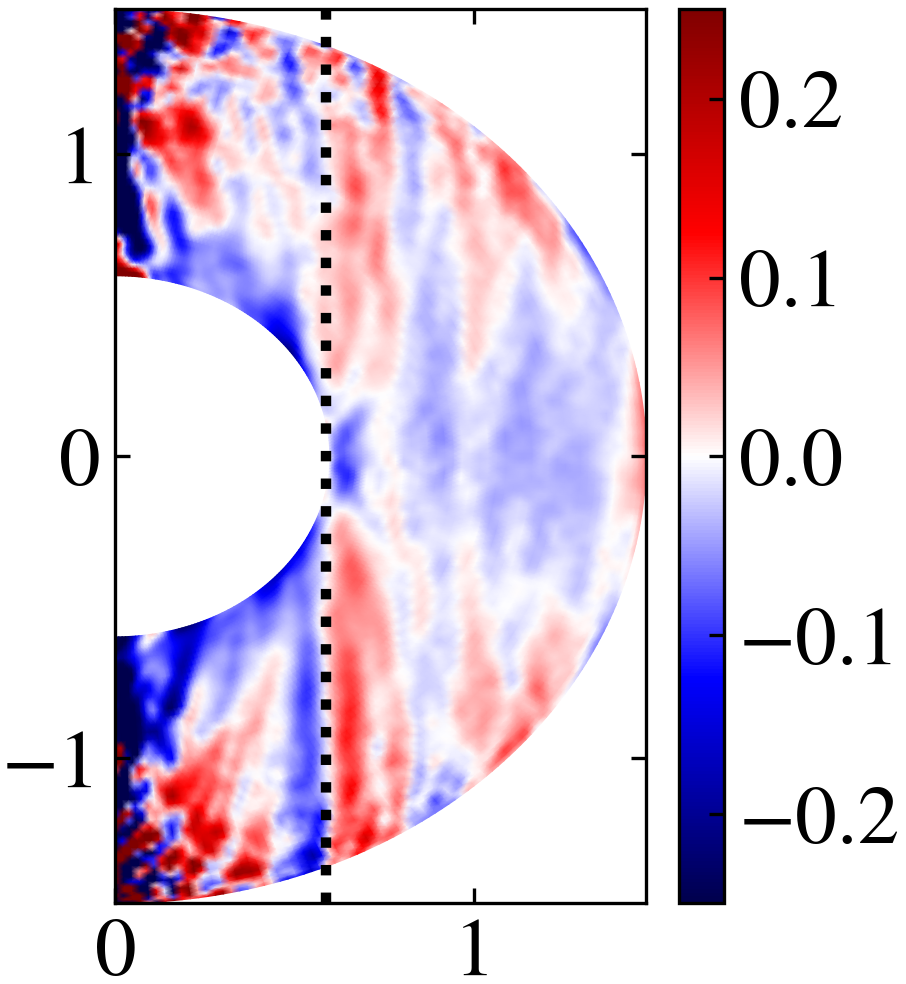}
         \caption{}
     \end{subfigure}
     \hfill
         \begin{subfigure}[b]{0.24\textwidth}
         \centering
         \includegraphics[width=\textwidth]{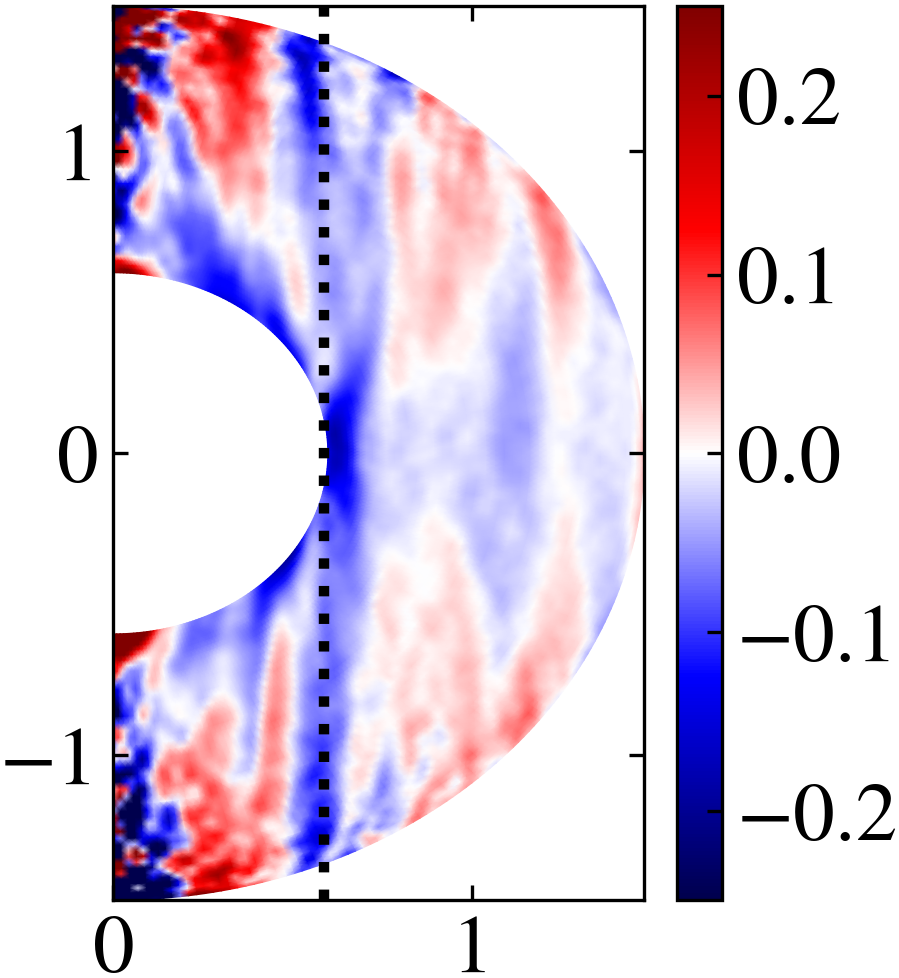}
         \caption{}
     \end{subfigure}
     \hfill
    \begin{subfigure}[b]{0.24\textwidth}
         \centering
         \includegraphics[width=\textwidth]{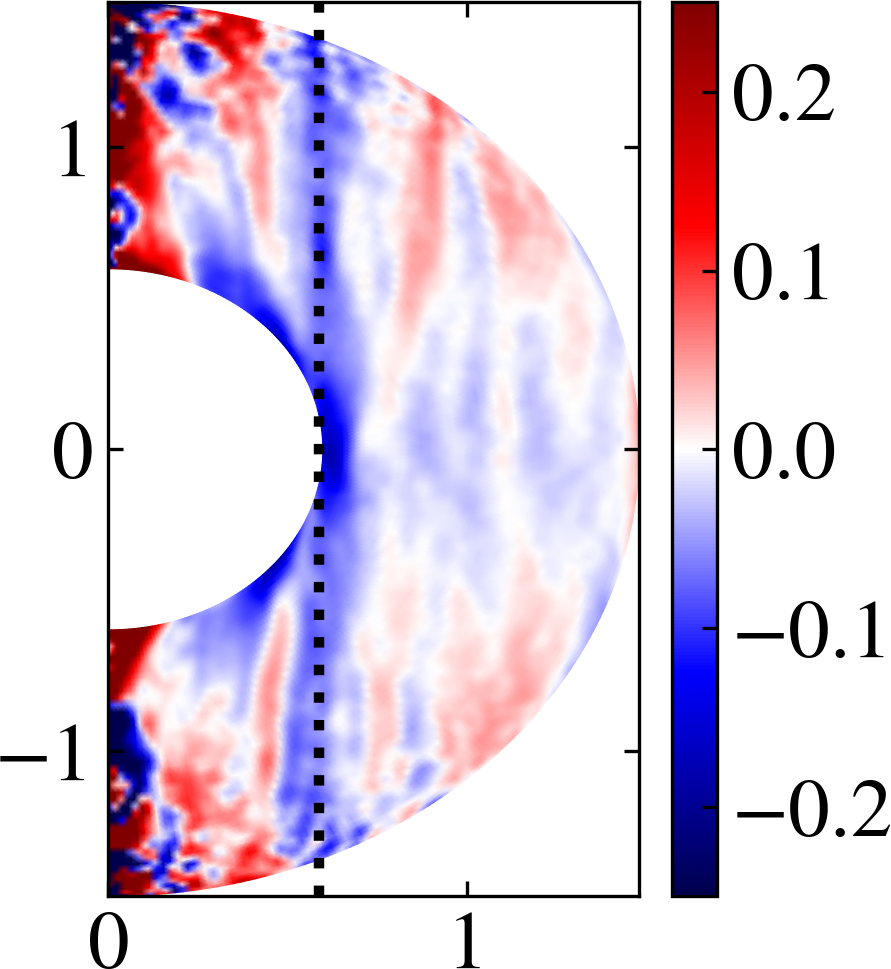}
         \caption{}
     \end{subfigure}
     \hfill
         \begin{subfigure}[b]{0.24\textwidth}
         \centering
         \includegraphics[width=\textwidth]{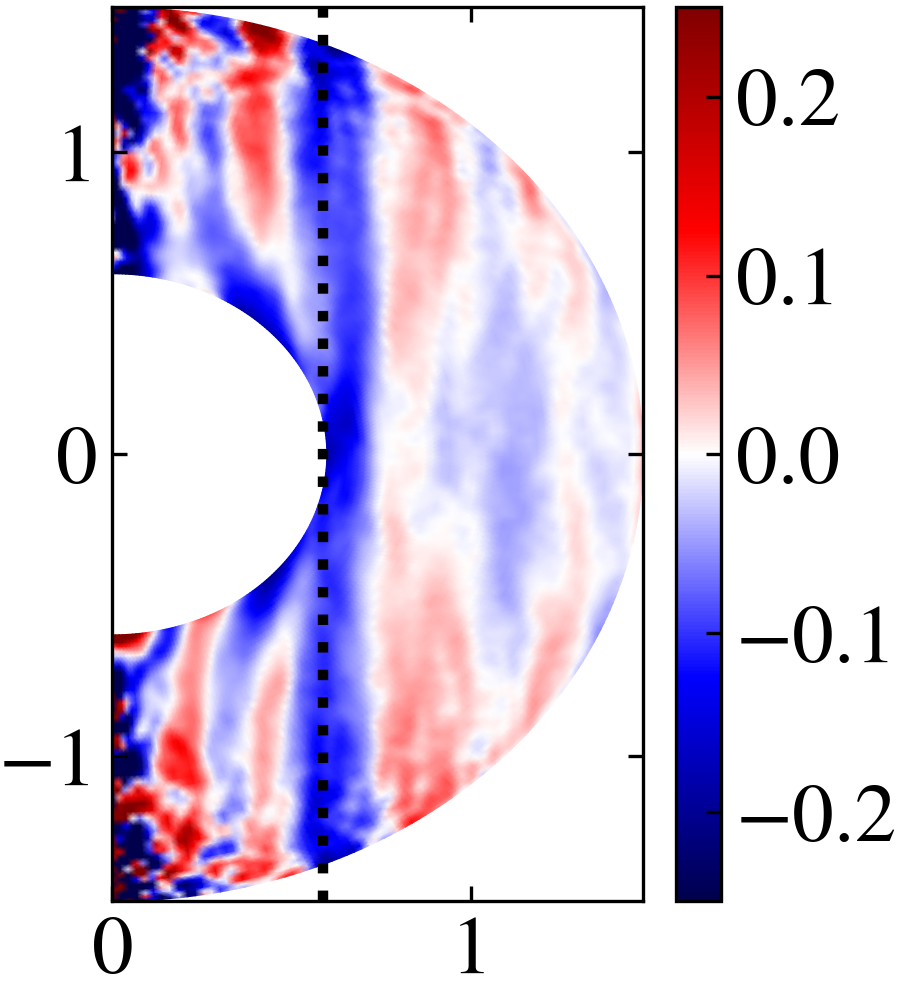}
         \caption{}
     \end{subfigure}
\caption{Distribution of \anna{dimensionless local radial shear $\Ross_{\rm sh}^l$ as  given in Eq.~(\ref{eq:rosh_rad_2d})} for the run with $g\sim 1/r^2$, $\Rayl= 5\times10^6$, $\xi = 0.35$. Dotted line denotes the location of tangent cylinder. (a) $\Ross_{\rm sh} = 0$,  \anna{without imposed differential rotation at the bottom of the convection zone. This case corresponds to naturally developing deviation from solid-body rotation with $\Omega_s$, due to the turbulent and magnetic stresses of the convective dynamo.}  (b) $\Ross_{\rm sh}=0.0125$, before collapse; (c) $\Ross_{\rm sh}=0.015$, at the dipole collapse; (d) $\Ross_{\rm sh}=0.02$, after the collapse. }
    \label{fig:Rosh_Roconv_local}
\end{figure*}

\begin{figure*}
    \centering
             \begin{subfigure}[b]{0.22\textwidth}
         \centering
         \includegraphics[width=\textwidth]{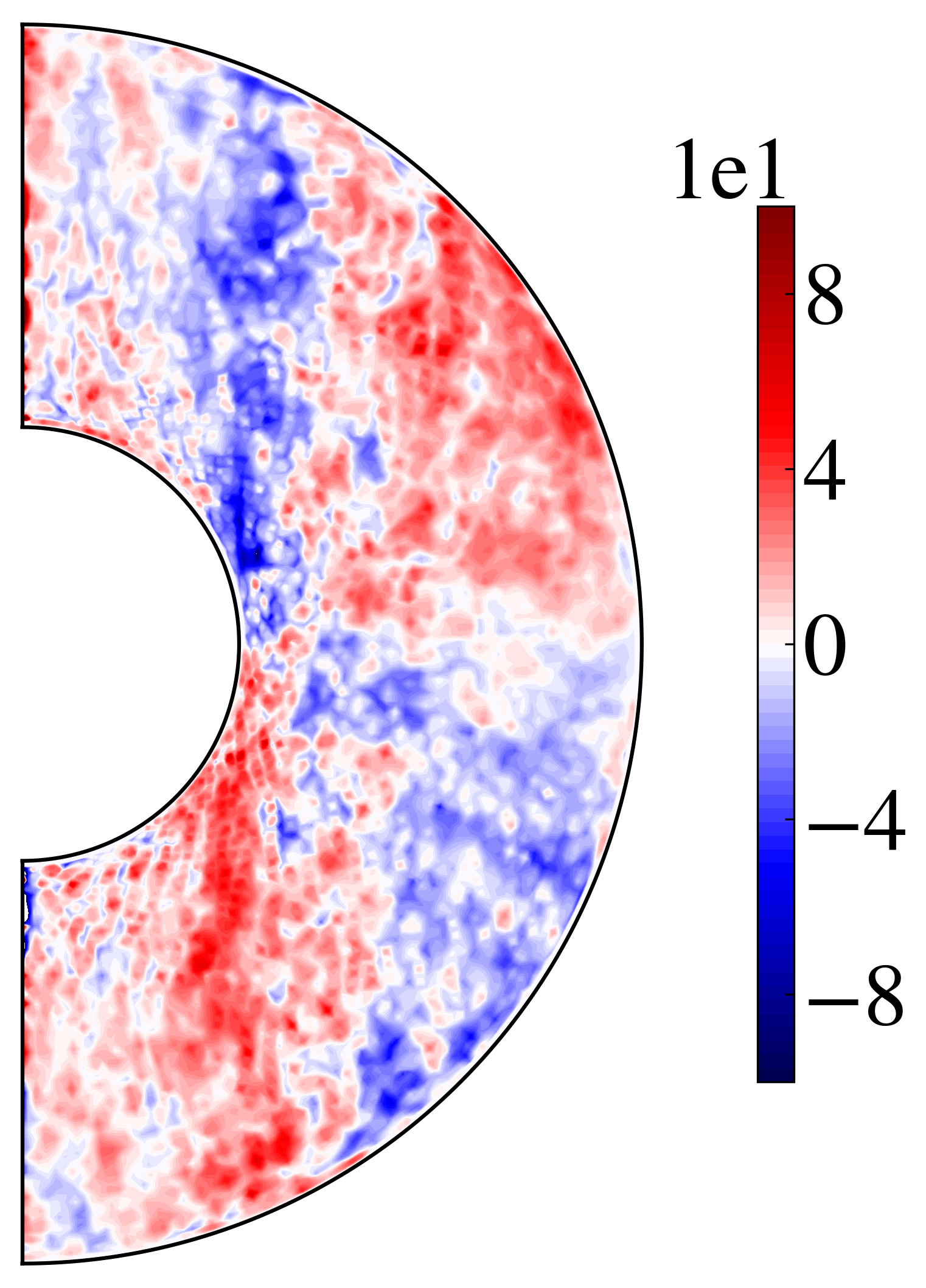}
         \caption{}
     \end{subfigure}
     \hfill
     \begin{subfigure}[b]{0.27\textwidth}
         \centering
         \includegraphics[width=\textwidth]{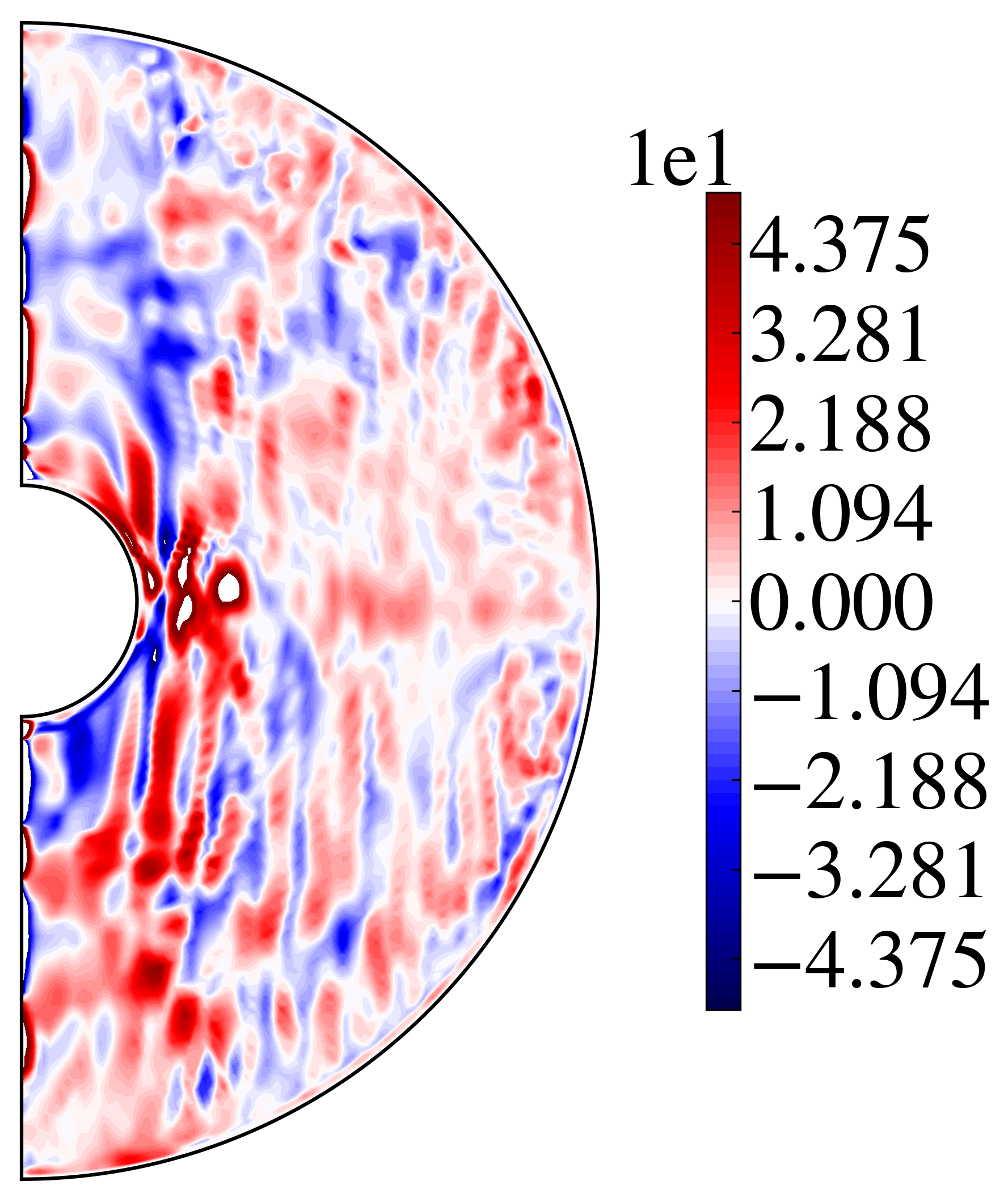}
         \caption{  }
     \end{subfigure}
     \hfill
          \begin{subfigure}[b]{0.22\textwidth}
         \centering
         \includegraphics[width=\textwidth]{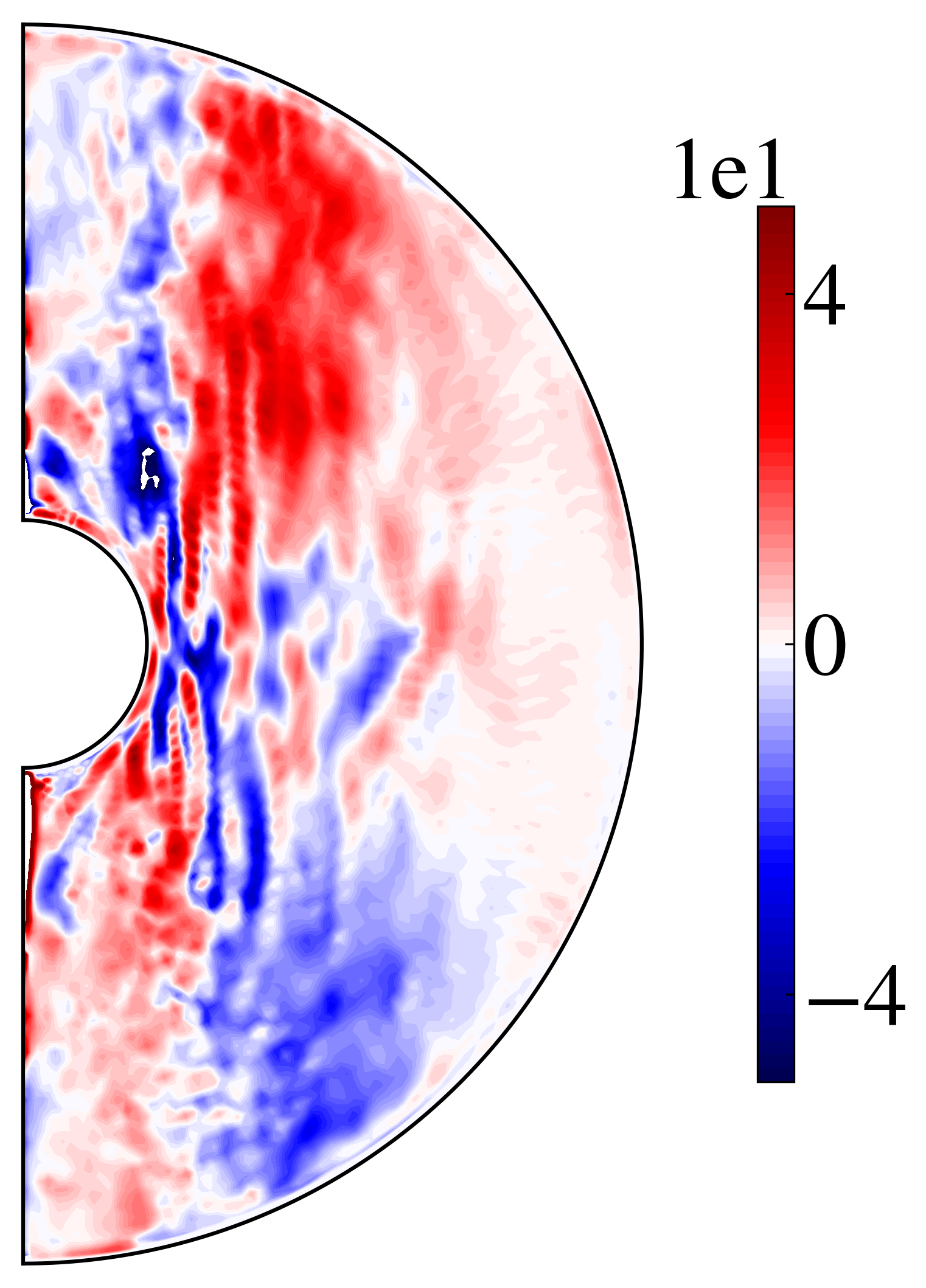}
         \caption{}
     \end{subfigure}
    \caption{The latitudinal component of the $\gamma$-effect, $\gamma_t$, corresponding to the three runs in Fig.~\ref{fig:slices}. (a) $g\sim/r^2$, $\Rayl=5\times10^6$, $\xi=0.35$; (b) $g \sim r$, $\Rayl=8\times10^6$, $\xi=0.2$; (c) $g \sim 1/r^2$, $\Rayl=1\times10^6$, $\xi=0.2$. The overall symmetry of $\gamma$-effect remains the same, despite the small-scale noise. \anna{Same units as in Fig.~\ref{fig:cesam}(c-e).}}
    \label{fig:gamma_t}
\end{figure*}

\begin{figure*}
    \centering
\begin{subfigure}[b]{0.47\textwidth}
         \centering
         \includegraphics[width=\textwidth]{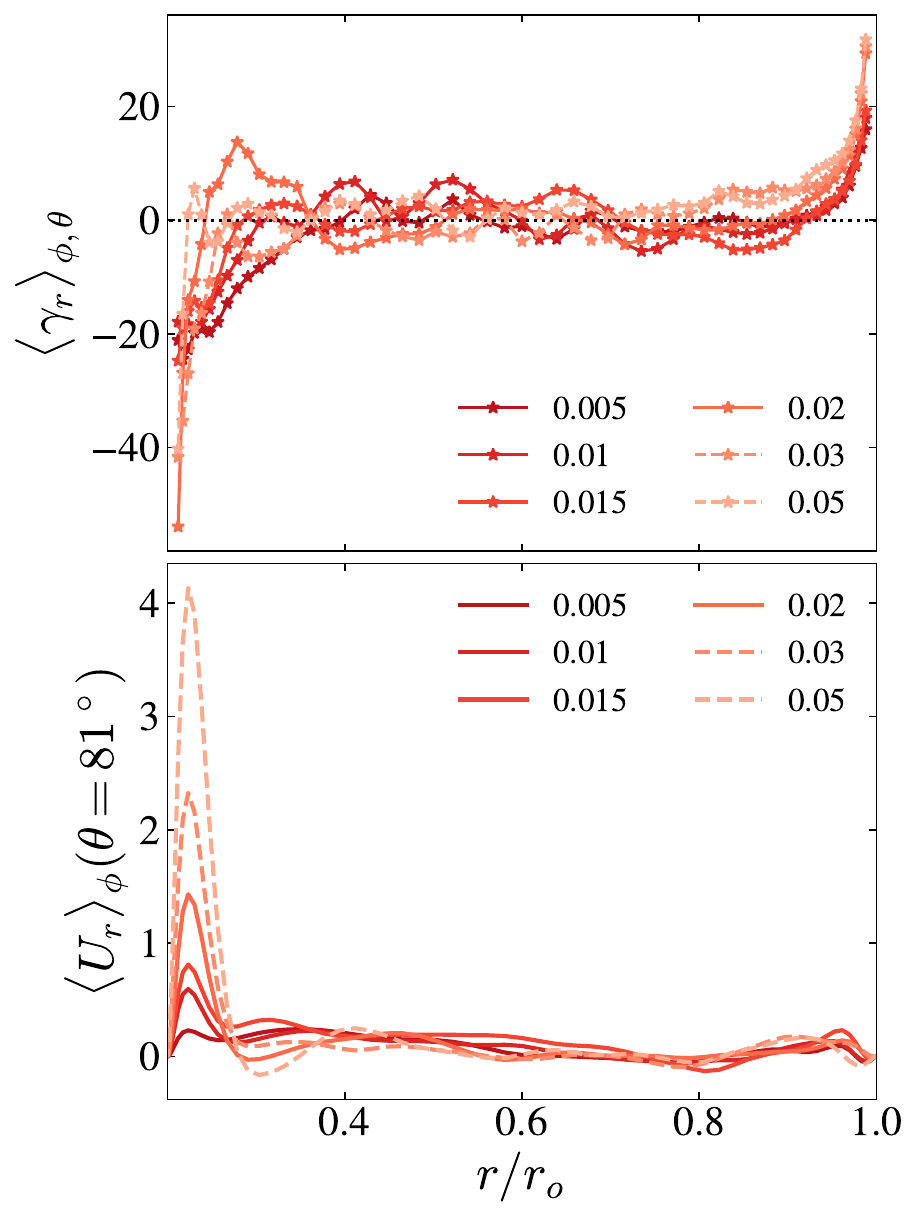}
         \caption{}
     \end{subfigure}
\hfill
     \begin{subfigure}[b]{0.47\textwidth}
         \centering
         \includegraphics[width=\textwidth]{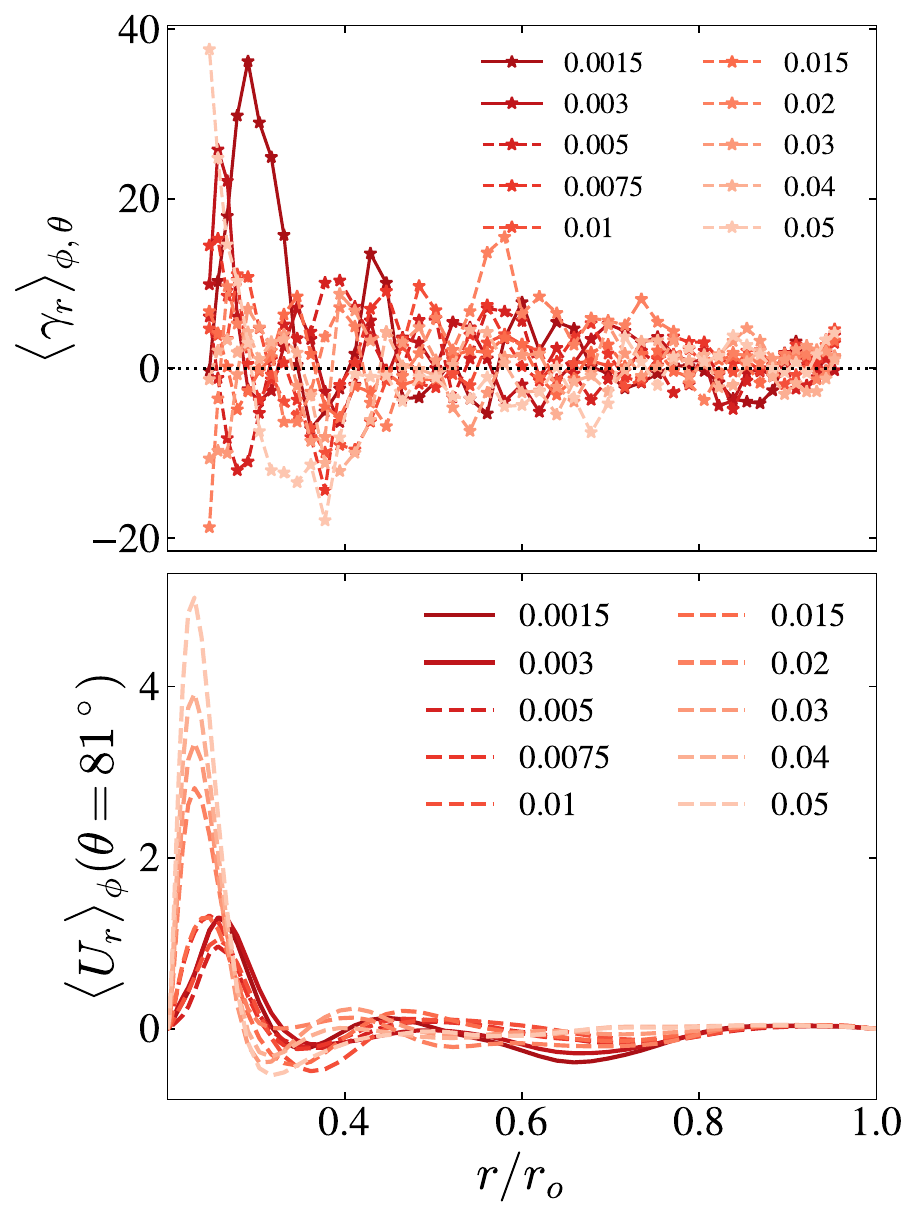}
         \caption{}
     \end{subfigure}
    \caption{Complementary to Fig.~\ref{fig:gamma_vs_Ur}. (a) Top: radial component of the gamma-effect $\gamma_r$ as a function of $Ro_{sh}$ (given in the legend), averaged over colatitude $\theta$.  Bottom: radial component of meridional circulation near the equator (colatitude of $\theta=80\circ$, see the dotted lines in Fig.~\ref{fig:slices}h,l). Solid lines indicate runs where the dipole is stable, and dashed lines to the runs where dipole collapsed. Runs with $g \propto r$, $\Rayl=8\times10^6$, $\xi=0.2$. (b) The same but for $g \propto 1/r^2$, $\Rayl=1\times10^6$, $\xi=0.2$. \anna{Subscripts $\phi$ and $\theta$ denote azimuthal and latitudinal average, respectively. Same units as in Fig.~\ref{fig:cesam}(c-e).} }
    \label{fig:gamma_vs_Ur2}
\end{figure*}

Here we present three additional figures, supporting findings of this work. Fig.~\ref{fig:Rosh_Roconv_local} presents a 2D distribution of local radial shear, according to Eq.~\ref{eq:rosh_rad_2d}, complementary to Fig.~\ref{fig:Rosh_Roconv_theta} where its distribution is compared at the midgap quantitatively for various values of $Ro_{\rm sh}$.  Fig.~\ref{fig:gamma_t} shows the latitudinal component of the $\gamma$-effect, $\gamma_t$, which together with the corresponding radial components $\gamma_r$ in Fig.~\ref{fig:slices} represents poloidal contribution to the mean-field advection of the large-scale magnetic field, according to~Eq.~\ref{eq:mean_field_B}. Fig.~\ref{fig:gamma_vs_Ur2} is complementary to Fig.~\ref{fig:gamma_vs_Ur} and depicts the competition between $\gamma_r$ and the radial component of poloidal circulation $\langle U_r \rangle_\phi$, for the two remaining cases of $g\propto r$ and $g\propto 1/r^2$, but with $\xi = 0.1$. To highlight the enhancement of radial meridional circulation with $Ro_{\rm sh}$, it is depicted at mid-latitudes of $\theta=\pi/3$. However, $\gamma$-effect, calculated with SVD method of~\cite{simard2016characterisation}, is too noisy in the area of the tangent cylinder. To highlight trends in radial distribution of $\gamma_r$ in Figs.~\ref{fig:gamma_vs_Ur2} and~\ref{fig:gamma_vs_Ur}, it was averaged over the latitude $\theta$.

 
\FloatBarrier

\FloatBarrier

\twocolumn

\section{Evolution of rotation properties in 1D models}
\label{app:prot_evol}



\FloatBarrier

\begin{figure}[h!]
    \centering
    \includegraphics[width=\linewidth]{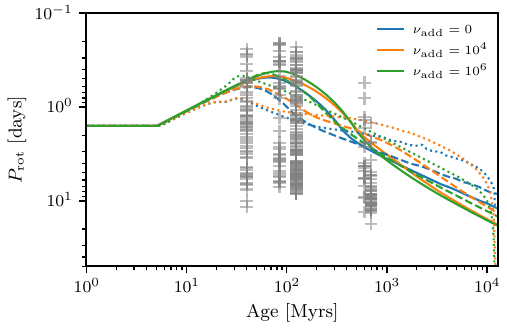}
    \caption{\lm{Surface rotation period as function of age for models with $0.55M_\odot$ (solid lines), $0.7M_\odot$ (dashed lines) and $1M_\odot$ (doted lines). The color corresponds to various values of additional viscosity (in unit of $\textrm{cm}^2~\textrm{s}^{-1}$). For all models, the initial conditions include a disk with a lifetime of 5 Myrs and a period of rotation of 10 days. Grey plus symbols correspond to the measured rotation period of stars in open clusters NGC2547 (40 Myrs), aPersei (85 Myrs), the Pleiades (125 Myrs), Praesepe (600 Myrs) and the Hyades (700 Myrs). Estimated properties taken from \citet{Wright2011}.}}
    \label{fig:prot_evol}
\end{figure}
\end{appendix}

\end{document}

%% file: intro_charly.tex
Pre-Main-Sequence (PMS) stars result from the gravitational collapse of molecular clouds and a subsequent accretion protostellar phase until the central hydrostatic core has mainly emerged from the dust circumstellar cocoon. Newly-born PMS stars then start their life as classical T-Tauri objects (CTTS) where they are still surrounded by a protoplanetary accretion disk before continuing in a non-accreting isolated contracting phase (e.g., WTTS, for weak line T-Tauri stars). As demonstrated by large sample X-ray and magnetospheric line measurements \citep[e.g.,][]{Argiroffi2016,Yamashita2020}, PMS stars are very active magnetic objects. Their activity level follows a relation with the surface convective Rossby number (i.e., representing the ratio $P_{\rm rot}/\tau_{\rm conv}$, with $P_{\rm rot}$ the rotation period and $\tau_{\rm conv}$ the typical convective timescale) very similar to what is observed in main-sequence stars. These observations suggest that the efficiency of stellar dynamo, the generation of magnetic fields by motions in their convective envelope, tends to drop along the PMS evolution \citep[e.g.,][]{getman2023magnetic}. Other evidence of such a transition are provided by spectro-polarimetric measurements, which permitted us to map the surface magnetic field in about thirty PMS stars \citep{Hill2017,Nicholson2021}. These observations indicate that younger and lighter stars are more likely to possess large-scale dipolar magnetic fields than older and heavier ones. Magnetic fields play an important role in the dynamics of PMS stars as they are involved in numerous processes: magnetospheric accretion and ejections, winds and associated braking torques, as well as internal angular momentum redistribution \citep[e.g.,][]{bouvier2014angular,Hartmann2016,Amard2023}. All these processes depend not only on the intensity but also on the topology of the magnetic fields \citep[e.g.,][]{Garraffo2016,Mueller2024}. Understanding these magnetic processes is crucial as the PMS evolution sets the conditions for planet formation (e.g., stellar activity, habitability) and future main-sequence evolution. 

Stellar magnetism and rotation are intrinsically interlaced, so that grasping the evolution of the former requires to model the latter. Observations of young clusters (ages of a few Myrs up to $\sim$20-30 Myrs) show a broad spread in rotation periods, ranging from very slow rotators (tens of days) to very fast ones \citep[less than few days; see e.g.,][]{scholz2007rotation,Moraux2013,Venuti2017}. This spread reflects both initial conditions and the interplay of spin-up and spin-down processes during the PMS phase \citep[e.g.,][]{bouvier2014angular}. On the one hand, young PMS T-Tauri stars possess strong magnetic fields (few kilogauss) that can interact with the circumstellar matter, constraining the near surface layers to co-rotate with the disk (disk locking phenomenon). Even when accretion stops (i.e., around an age of few Myrs), stellar winds still continue to remove angular momentum from the stellar surface. On the other hand, in deep layers, a contracting radiative core emerges on the PMS phase, pushing progressively its border with the convective envelope up \citep[e.g.,][]{Kunitomo2018}. As a result of the interplay between the inner collapse of the radiative core and the surface magnetic braking, the core-to-envelope rotation rate ratio is expected to increase up to one order of magnitude \citep[e.g.,][]{Gallet2013}. This radial differential rotation is very likely to play a role in shaping the convective cells in the envelope of PMS stars and their large-scale magnetic properties, especially through the dynamo $\Omega$-effect \citep{Moffat78}.

Modeling the self-consistent generation of magnetic fields by density-stratified turbulent rotating convection is a very complex task from a theoretical point of view, as it involves very different scales interacting non-linearly and three-dimensionally \citep[e.g.,][]{Kapyla2023}. Previous numerical work relied on the anelastic formulation of the MHD equations that permit to focus on the convective dynamics while filtering (fast) acoustic waves \citep[e.g.][]{jones2011anelastic}. Such direct numerical simulations (DNS) of rotating spherical convective shells exhibit a wide variety of dynamo states: mechanisms at work are complex and the simple question of knowing whether the dipolar component of the magnetic field dominates or not still remains valuable, even if clues have already been obtained through parameter studies \citep[e.g.][]{Raynaud2015,Zaire2022,Guseva2025}. Previous investigation showed in particular that self-consistently developed differential rotation tends to decrease the relative dipole strength and drives magnetic cycles through so-called oscillatory dynamos \citep{parker1955hydromagnetic,yoshimura1975solar,schrinner2012dipole}. Few recent simulations also accounted for the effect of a radiative core below the convective envelope, allowing for the presence of a tachocline \citep[e.g,][]{Brun2022}, but to our knowledge, none of them systematically studied the effect of a high contraction-induced inner spin-up.

In this work, we aim to elucidate how a large-scale radial differential rotation expected in the convective zones of PMS stars can affect the magnetic field generation and, in particular, the stability of dipolar dynamos. To do so, we perform new 3D simulations of thick rotating spherical convective envelopes with a varying imposed shear between the bottom and top boundaries, extending previous spherical Couette-flow simulations in stably-stratified regions \citep[e.g.,][]{Guervilly2010,Petitdemange2024} to convective regions. This setup mimics at lower cost the expected radial shear induced through the combined effect of the radiative core contraction and the different sources of surface braking in PMS stars, and represents a first step in the investigation of such an effect. A parameter study then enables us to establish a criterion for collapse of large-scale dipole field suited for 1D stellar structure models, and ultimately, to discuss the origin of the magnetic properties distribution in the mass-period diagram on the PMS.

The paper is organized as follows. In Sect.~\ref{section:early_stell_evol}, we recall the main structural properties of PMS stars in the light of the stellar evolution code \cesamxx. This preliminary section guides the physical setup considered in our MHD dynamo simulations. The outcomes of the simulations are presented in Sect. \ref{section:dipole_stability}. In particular, we propose a local and global criteria for the dipole stability based on the level of imposed radial differential rotation. The mechanisms for dipole collapse in the presence of differential rotation are described in detail in Sect. \ref{section:mechanism_collapse}. In Sect. \ref{section:discussion}, the results are discussed. As a first application, we analyze the predictions of the latter criterion on grids of 1D stellar models and confront the results to observations. We conclude in Sect.~\ref{section:concl}.

%% file: 32_charly.tex
In a preliminary step, it is worth discussing the impact of the stellar gravity profile evolution on the convective flow properties. First, in the limiting case of $g\propto r$, as expected at very early stages of stellar formation (i.e., quasi uniform density), the convection seems more vigorous near the outer shell, as shown in Fig.~\ref{fig:cesam}d. For the same values of $N_\rho=2.5$ and $\xi=0.2$, assuming this time $g\propto 1/r^2$, as expected for instance at the end of the PMS stage (i.e., matter is concentrated in the inner core), convective columns become very concentrated around the bottom boundary, as shown in Fig.~\ref{fig:cesam}e. Unsurprisingly, convective velocities are higher where gravity (and thus buoyancy driving) is maximum for such a relatively mild density contrast. We note that larger density contrasts would counteract the effect of a decreasing gravity profile and also favor convective motions in the very outer shells, similarly to a gravity profile $g\propto r$ \citep[e.g.][]{GastineDW12,Raynaud2014,Guseva2025}. For a thinner convective shell (i.e., $\xi=0.35$), we see in Fig.~\ref{fig:cesam}c that the convective velocity is more homogeneously distributed in the convective bulk.

Dynamo action is promoted by helicity, which is directly correlated with the convective Rossby number. As seen in Eq.~\ref{eq:roc}, the latter is proportional to the inverse of the convective characteristic length scale. In Fig.~\ref{fig:cesam}b, we plot the radial profiles of this quantity for different gravity profiles. In case $g\propto r$, $\Ross_{\rm conv} (r)$ has a clear maximum in the outer layers where the convective velocity is large and associated with small scales. In case $g\propto 1/r^2$, $\Ross_{\rm conv} (r)$ is also maximum close to the surface, although more homogeneously distributed than in case $g\propto r$. This statement holds true even in case $\xi=0.2$, despite a maximum velocity close to the center, as seen in Fig.~\ref{fig:cesam}e. For a gravity profile comparable to that of \cesamxx PMS models, the $\Ross_{\rm conv} (r)$ profile remains between the two previous limiting cases. This systematic increasing trend towards the surface mainly results from the sharp drop of density close the top boundary; this promotes very small scale motions, and thus high values of $\Ross_{\rm conv} (r)$ and helicity. As a result, we see that the gravity profile only slightly affects the radial distribution of $\Ross_{\rm conv} (r)$ and thus helicity. 
Because differences are small, dynamo mechanisms of large-scale field generation are thus expected to be universal for all gravity profiles, as we see in Sect.~\ref{section:mechanism_collapse}.